\pdfoutput=1
\documentclass[aps,prd,amsmath,floats,floatfix,twocolumn,superscriptaddress,nofootinbib,showpacs,longbibliography]{revtex4-1}

\usepackage[T1]{fontenc}
\usepackage[utf8]{inputenc}
\usepackage[english]{babel}
\usepackage{txfonts}
\usepackage{microtype}
\usepackage{amssymb,bm,mathrsfs}
\usepackage[nomessages]{fp}
\DeclareMathAlphabet{\mathpzc}{OT1}{pzc}{m}{it}
\usepackage{graphicx}
\usepackage{subfigure}
\usepackage{array}[=2016-10-06] % compatibility with REVTeX
\usepackage{booktabs}
\usepackage{verbatim}
\usepackage{xspace}
\usepackage[normalem]{ulem}
\usepackage[dvipsnames,table]{xcolor}
\definecolor{linkcolor}{rgb}{0.0,0.3,0.5}
\usepackage[hypertexnames=false,unicode,colorlinks=true,linkcolor=linkcolor,citecolor=linkcolor,filecolor=linkcolor,urlcolor=linkcolor,pdfusetitle]{hyperref}
\defcitealias{Islam:2025drw}{IW25}

\begin{document}
\title{\texttt{gwBenchmarks}: Evaluating scientific agents in high-precision gravitational-wave astronomy}

\author{Tousif Islam}
\email{tousifislam@ucsb.edu}
\affiliation{Kavli Institute for Theoretical Physics,University of California Santa Barbara, Kohn Hall, Lagoon Rd, Santa Barbara, CA 93106}
\author{Digvijay Wadekar}
\affiliation{\mbox{Center for Gravitational Physics, University of Texas at Austin, Austin, TX 78712, USA}}
\affiliation{\mbox{Department of Physics and Astronomy, Johns Hopkins University, 3400 N. Charles Street, Baltimore, Maryland, 21218, USA}}
\author{Zihan Zhou}
\affiliation{\mbox{Department of Physics, Princeton University, Princeton, NJ 08540, USA}}

%%%%%%%%%%%%%%%%%%%%%%%%%%%%%%%%%%%%%%%%%%%%%%%%%%%%%%%%%%%%%%%%%%%%%%%%%%%
% Because hyperref only gets the *last* author, we need to be explicit.
\hypersetup{pdfauthor={Islam et al.}}
\date{\today}

%==========================================================================
%==========================================================================
%==========================================================================
%==========================================================================
\begin{abstract}
Interactive scientific agents can produce executable code and favorable
self-reported scores while failing the underlying scientific task. We propose evaluating such agents at three complementary levels: \emph{integrity} of the submission, \emph{outcome} under an independent scientific metric, and \emph{diagnosis} of the recorded trajectory. We apply this framework to \texttt{gwBenchmarks}, eight end-to-end tasks from
gravitational-wave astronomy grounded in high-fidelity simulations and catalogs representing over $5\times10^8$ core-hours of computation. To prevent an agent's evaluation errors from determining its score, we separate artifact generation from scoring: a separate evaluator checks whether submissions follow the protocol and calculates physics-grounded metrics on complete validation sets.
Across fifteen agent configurations, two fail integrity checks before outcome scoring, while the remaining thirteen expose partial validation, metric substitution, constraint violations, and fabricated outputs. No scored agent consistently dominates across tasks and the performance remains significantly below the physics target in a few cases. We also use fixed python scripts to examine saved code, scorecards, reports, and changelogs and find evidence of failures across task understanding, method selection, code, evaluation, physics, verification, and reporting. Together, these results show why evaluation of interactive scientific agents must combine submission integrity, independently verified scientific outcomes, and diagnosis of the recorded trajectory.
\end{abstract}
%==========================================================================
%==========================================================================

\maketitle

%==============================================================================================
%==============================================================================================
\section{Introduction}
\label{sec:introduction}
%==============================================================================================
%==============================================================================================

Scientific agents are increasingly capable of carrying out extended computational workflows that resemble parts of scientific research: interpreting a problem, selecting methods, writing and executing code, inspecting numerical results, constructing models, and revising their approach. Evaluating such systems requires substantially more than checking whether a task terminates or the resulting code executes. An agent can complete an apparently coherent workflow and produce a plausible scientific artifact while using the wrong metric, evaluating only an easy subset of the data, violating a physical constraint, or reporting a result that cannot be reproduced. Conversely, an agent may construct a scientifically interesting model even when its final quantitative accuracy remains inadequate. Evaluating scientific agents therefore requires both challenging end-to-end benchmarks and verification procedures that determine whether the resulting artifacts are actually scientifically valid.

Gravitational-wave (GW) physics provides a particularly demanding setting for such evaluation. Modern GW modeling combines expensive numerical simulations, perturbative and analytical calculations, surrogate modeling, regression, time-series prediction, numerical optimization, and detector-specific validation~\citep{loeffler2011einsteintoolkit,blackman2017precessingsurrogate,varma2019precessingsurrogate,wette2020swiglal,vandenbroeck2009templatebanks}. Scientific adequacy is often determined by precise quantitative metrics rather than by qualitative plausibility or successful code execution. Small modeling errors can be consequential for waveform analysis even when the corresponding numerical output appears visually reasonable. At the same time, the field provides public high-fidelity simulations and well-defined physical observables that make many scientific artifacts independently testable.

We therefore introduce \texttt{gwBenchmarks}$^{\ref{data_link}}$, an end-to-end benchmark for scientific agents in gravitational-wave physics. To our knowledge, it is the first dedicated benchmark for agentic scientific modeling in this domain. The suite contains eight heterogeneous research tasks spanning interpolation, regression, high-dimensional time-series modeling, constrained analytic modeling, scientific optimization, and physics derivation and implementation. Agents interact with a computational environment by inspecting data, writing and executing code, testing candidate methods, constructing scientific artifacts, and revising their solutions. The benchmark source simulations and catalogs represent more than $5\times10^8$ core-hours of computation, while several tasks demand accuracies sufficiently stringent that scientifically consequential failures produce no obvious execution error. The suite therefore tests not merely whether an agent can generate scientific code, but whether it can construct quantitatively useful scientific artifacts through an extended research workflow.

Difficult end-to-end scientific benchmarks introduce a second problem: the agent may implement not only the scientific artifact but also the procedure used to evaluate it. A favorable self-reported score can therefore reflect an incorrect metric, incomplete validation, or an invalid scientific assumption rather than a successful result. We address this problem with a complementary evaluation framework that separates artifact generation from scientific verification. It asks three questions. First, \emph{integrity}: do the submission and its accompanying scientific record satisfy the task protocol? Second, \emph{outcome}: how well does the submitted artifact perform when a separate evaluator applies the prescribed scientific metric to the complete validation set? Third, \emph{diagnosis}: what auditable evidence in the saved code, scorecards, reports, and changelogs reveals problems in task understanding, method selection, implementation, evaluation, physics, verification, or reporting? Together, these levels distinguish invalid or misleading submissions from valid but scientifically inadequate artifacts, while also providing evidence about how those outcomes arose.

\begin{table*}[t]
\centering
\caption{\textbf{Positioning of representative benchmark paradigms relevant to scientific agents.} The columns distinguish requirements that become important when agents construct scientific artifacts through extended computational workflows. ``Partial'' indicates that a property is present for some tasks or aspects of an evaluation rather than being a defining requirement of the benchmark.}
\label{tab:benchmark_comparison}
\small
\begin{tabular}{lcccc}
\toprule
\textbf{Benchmark} &
\shortstack{\textbf{Long-}\\\textbf{horizon}} &
\shortstack{\textbf{Scientific}\\\textbf{validity}} &
\shortstack{\textbf{Standardized}\\\textbf{evaluation}} &
\shortstack{\textbf{Artifact}\\\textbf{verification}} \\
\midrule
SWE-bench & \checkmark & $\times$ & Partial & $\times$ \\
ScienceAgentBench & Partial & Partial & $\times$ & $\times$ \\
PDEBench & $\times$ & \checkmark & \checkmark & $\times$ \\
\texttt{gwBenchmarks} & \checkmark & \checkmark & \checkmark & \checkmark \\
\bottomrule
\end{tabular}
\end{table*}

We attempt the eight-task suite with fifteen LLM coding-agent configurations. Two configurations fail integrity requirements before outcome comparison, leaving thirteen scored configurations. For the remaining submissions, a separate evaluation process applies fixed task-specific metrics to complete validation sets rather than adopting agent-reported scores~\citep{austin2021programsynthesis,jimenez2023swebench,chen2024scienceagentbench,merrill2026terminalbench}. Performance varies substantially across scientific tasks. Agents solve several low-dimensional interpolation and regression problems to high accuracy, and in some cases independently construct physically meaningful representations, including a closed-form inspiral--merger--ringdown waveform model and a coordinate transformation that regularizes the near-extremal behavior of Kerr quasi-normal-mode frequencies. More demanding modeling tasks remain substantially harder: for example, the best waveform surrogate has a mismatch roughly $30\times$ above the target accuracy. At the same time, we observe partial validation, metric substitution, constraint violations, numerical instability, fabricated outputs, and other failures that would be obscured by execution success or agent-reported scores alone.

The contributions of this work are therefore twofold. First, we introduce \texttt{gwBenchmarks}, a heterogeneous end-to-end benchmark for evaluating scientific agents on quantitatively demanding tasks in gravitational-wave physics. The benchmark probes not only code generation but the construction of scientific artifacts across multiple forms of modeling, optimization, and theoretical implementation. Second, we introduce and apply a three-level verification framework based on submission \emph{integrity}, independently measured scientific \emph{outcome}, and \emph{diagnosis} from auditable records of the agent's work. Together, the benchmark and evaluation framework provide a controlled setting for studying both what current scientific agents can construct and how reliably their apparent successes can be verified. We release the benchmark, evaluation pipeline, agent outputs, and per-sample results as an open resource. Our \href{https://github.com/tousifislam/gwBenchmarks}{code}\footnote{\label{code_link}\href{https://github.com/tousifislam/gwBenchmarks}{https://github.com/tousifislam/gwBenchmarks}}, \href{https://huggingface.co/datasets/GWagents/gwBenchmarks}{data}\footnote{\label{data_link}\href{https://huggingface.co/datasets/GWagents/gwBenchmarks}{https://huggingface.co/datasets/GWagents/gwBenchmarks}}, and \href{https://tousifislam.com/gwBenchmarks/}{website}\footnote{\label{web_link}\href{https://tousifislam.com/gwBenchmarks/}{https://tousifislam.com/gwBenchmarks/}} are publicly available.

%==============================================================================================
%==============================================================================================
\section{Related Work and Benchmark Positioning}
\label{sec:relatedworks}
%==============================================================================================
%==============================================================================================

Language-model evaluation has progressed from knowledge and reasoning suites such as \texttt{MMLU}, \texttt{BIG-bench}, and \texttt{HELM} to executable coding and repository-level tasks~\citep{hendrycks2020mmlu,srivastava2022bigbench,liang2023helm,chen2021evaluatingcodellms,jimenez2023swebench,merrill2026terminalbench}. Interactive agent benchmarks further evaluate reasoning--action loops, tool use, and navigation through simulated or computational environments~\citep{yao2023react,liu2023agentbench,mialon2023gaia,zhou2023webarena}. These benchmarks primarily establish general reasoning, coding, and interactive capabilities rather than domain-specific scientific validity.

Scientific machine-learning benchmarks instead emphasize quantitative prediction in particular scientific domains~\citep{dunn2020matbench,wu2017moleculenet,chanussot2021opencatalyst,takamoto2022pdebench,watsonparris2021climatebench,huang2025cosmobench}. More recent benchmarks increasingly evaluate agents carrying out scientific workflows. \texttt{ScienceAgentBench} uses executable programs and expert-validated tasks drawn from published research; \texttt{BLADE} evaluates fine-grained analytical decisions; and \texttt{DiscoveryBench} evaluates components of multi-step scientific discovery~\citep{chen2024scienceagentbench,gu2024blade,majumder2024discoverybench}. In physics, \texttt{NewtonBench} and \texttt{DiscoverPhysics} evaluate interactive law discovery in simulated systems~\citep{zheng2025newtonbench,wiemann2026discoverphysics}.

These benchmark families emphasize complementary aspects of scientific AI. Coding benchmarks test whether agents can construct functioning software; scientific machine-learning benchmarks provide standardized quantitative prediction tasks; and scientific-agent benchmarks test increasingly open-ended analysis, experimentation, and discovery. \texttt{gwBenchmarks} targets an intersection that has received less attention: extended agentic workflows that produce heterogeneous scientific artifacts whose validity and accuracy can be quantitatively tested in a high-precision physical domain. Table~\ref{tab:benchmark_comparison} positions representative benchmarks according to several aspects particularly relevant to this setting.

Here, \emph{long-horizon} denotes multi-step computational interaction involving activities such as implementation, execution, debugging, and revision; \emph{scientific validity} denotes evaluation against domain-specific numerical or physical requirements; \emph{standardized evaluation} denotes performance determined through a pre-specified procedure rather than an agent's own assessment; and \emph{artifact verification} denotes explicit checks that the submitted scientific object satisfies required constraints and can be validly evaluated. These dimensions are not intended as a general ranking of benchmarks: different benchmarks are designed to measure different capabilities.

Within this landscape, \texttt{gwBenchmarks} combines end-to-end artifact construction with quantitative, task-specific verification. Its distinction is not simply that the tasks are scientific, but that agents must construct multiple kinds of scientific object---including regression models, high-dimensional time-series surrogates, closed-form analytic expressions, optimized template banks, and implementations derived from theoretical calculations---under domain-specific accuracy and validity requirements. This positioning complements benchmarks centered on software correctness, static prediction, or simulated discovery, and provides a setting in which scientific capability and the reliability of its evaluation can be studied together.

%==============================================================================================
%==============================================================================================
\section{\texttt{gwBenchmarks}: Eight End-to-End Gravitational-Wave Tasks}
\label{sec:gwbenchmarks}
%==============================================================================================
%==============================================================================================

\texttt{gwBenchmarks} comprises eight end-to-end scientific tasks in GW physics. To aid readers unfamiliar with GW terminology, we provide a dedicated glossary in Appendix~\ref{sec:glossary}. The tasks draw on distinct components of the GW modeling workflow: numerical-relativity waveform approximation, black-hole remnant prediction, orbital dynamics, Kerr ringdown, waveform-validity estimation, analytic waveform construction, template-bank optimization, and the implementation of new theoretical corrections. Together, they probe regression, interpolation, high-dimensional time-series and surrogate modeling, closed-form analytic modeling, scientific optimization, and physics derivation and implementation. This heterogeneity is deliberate: success at interpolation or low-dimensional regression need not imply success at high-dimensional waveform modeling, constrained analytic construction, optimization, or derivation-to-implementation reasoning. Despite their different scientific objectives, every task requires the agent to transform a specified problem and supplied reference data into a concrete, quantitatively testable artifact.

At an abstract level, the tasks share the same scientific workflow:
\begin{equation}
\underbrace{\textit{Reason} \rightarrow \textit{Code} \rightarrow \textit{Build}}_{\textit{recorded trajectory}} \rightarrow \underbrace{\textit{Submission}}_{\textit{integrity}} \rightarrow \underbrace{\textit{Prescribed metric}}_{\textit{outcome}}.
\end{equation}
The recorded work, submitted artifact, and independently evaluated outcome therefore provide complementary evidence about agent performance. What changes across the eight tasks is the scientific content and where the principal difficulty lies: constructing an accurate representation, respecting physical or methodological constraints, performing an optimization, translating theoretical expressions into executable software, or correctly establishing the accuracy of the final artifact. This diversity allows us to examine whether successful execution, protocol compliance, independently measured scientific performance, and diagnostic evidence from the work record tell a consistent story.

\begin{figure*}[t]
\centering
\includegraphics[width=0.88\linewidth]{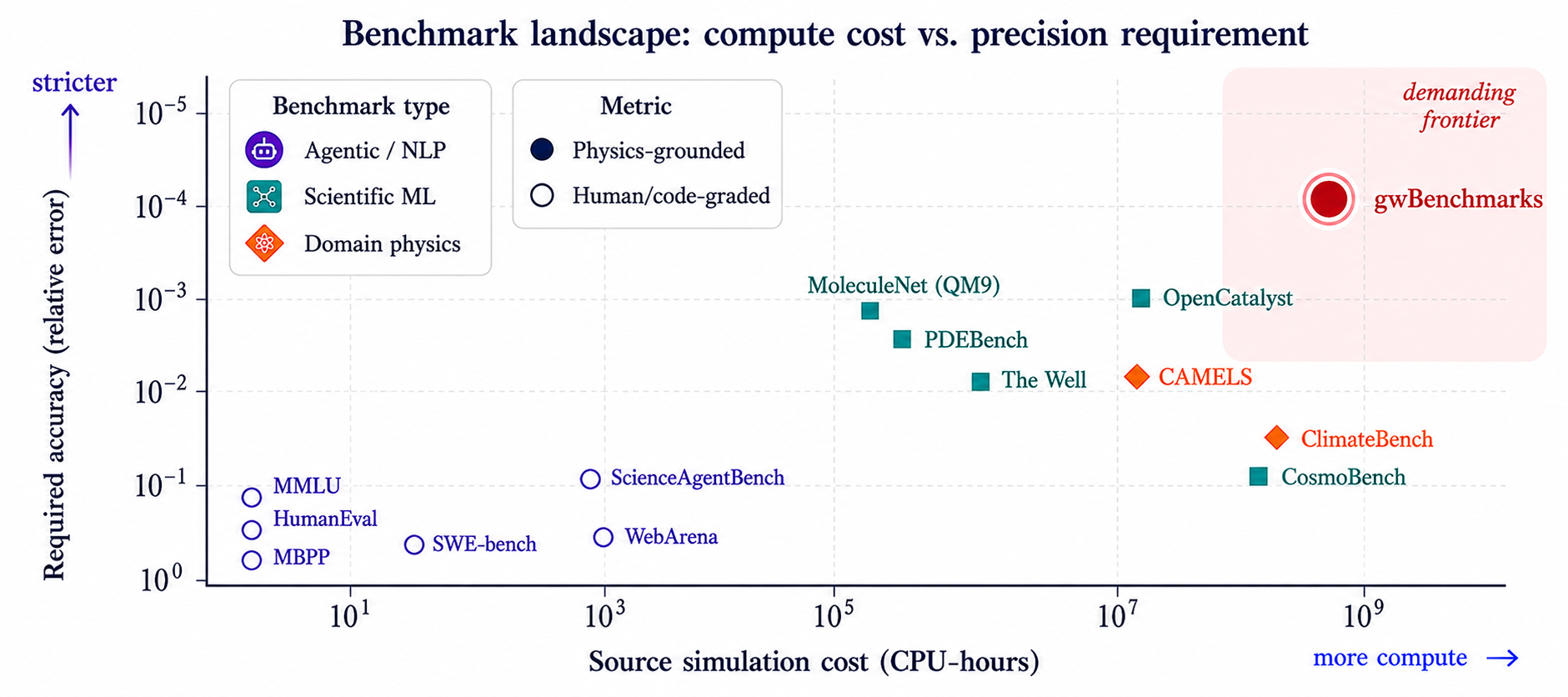}
\caption{
\textbf{\texttt{gwBenchmarks} targets end-to-end scientific modeling in a high-cost, high-precision regime.}
The x-axis shows the computational cost represented by the source simulations, while the y-axis shows the required relative accuracy, with lower values indicating stricter tolerances. Marker shape distinguishes benchmark type and fill distinguishes human/code-graded from physics-grounded evaluation. The combination of expensive reference calculations and stringent accuracy requirements motivates quantitative verification of agent-produced scientific artifacts rather than reliance on execution success or self-reported scores.
}
\label{fig:verification_regime}
\end{figure*}

Figure~\ref{fig:verification_regime} places the benchmark in the regime of source-data cost and required accuracy relevant to these tasks. The purpose is not to define difficulty through computational cost alone. Rather, the figure highlights the combination of expensive reference calculations and stringent downstream accuracy requirements: an artifact can execute successfully and appear numerically plausible while remaining inadequate for its intended scientific use.

Table~\ref{tab:benchmark_summary} provides the main-text overview of all eight tasks, summarizing for each the scientific objective, required agent-produced artifact, canonical evaluation quantity, and principal scientific challenge. The first five tasks primarily probe prediction, interpolation, regression, and time-series modeling, whereas the final three emphasize constrained analytic construction, scientific optimization, and derivation-to-implementation reasoning. Detailed descriptions of the data, parameter domains, physical assumptions, task constraints, and required outputs are provided in Appendix~\ref{sec:individual_benchmarks}. Exact definitions of the evaluation metrics, normalization conventions, and validation procedures are given in Appendix~\ref{app:metrics}. A glossary of GW terminology is provided in Appendix~\ref{sec:glossary}.

\begin{table*}[t]
\centering
\caption{
\textbf{Scientific tasks used to evaluate interactive agents in \texttt{gwBenchmarks}.}
The benchmark comprises eight end-to-end tasks spanning interpolation and regression, high-dimensional time-series and surrogate modeling, closed-form analytic construction, scientific optimization, and physics derivation and implementation. Each task requires an agent to construct a scientific artifact, with the scientific objective and canonical evaluation metric specified in advance and applied independently to the submitted artifact. The columns summarize the scientific objective, required agent-produced artifact, canonical evaluation quantity, and primary evaluation challenge for each task. Section~\ref{sec:gwbenchmarks} describes the benchmark design and scientific scope. Detailed task definitions, including the underlying data, parameter domains, physical assumptions, constraints, and required outputs, are provided in Appendix~\ref{sec:individual_benchmarks}; exact definitions of the evaluation metrics, normalization conventions, and validation procedures are given in Appendix~\ref{app:metrics}. A glossary of GW terminology is provided in Appendix~\ref{sec:glossary}.
}
\label{tab:benchmark_summary}

\scriptsize
\setlength{\tabcolsep}{3pt}
\renewcommand{\arraystretch}{1.12}

\begin{tabular}{@{}lllll@{}}
\toprule
\parbox[t]{0.10\textwidth}{\textbf{Benchmark}} &
\parbox[t]{0.18\textwidth}{\textbf{Scientific objective}} &
\parbox[t]{0.20\textwidth}{\textbf{Agent-produced artifact}} &
\parbox[t]{0.15\textwidth}{\textbf{Canonical evaluation}} &
\parbox[t]{0.28\textwidth}{\textbf{Primary evaluation challenge}} \\
\midrule

\parbox[t]{0.10\textwidth}{\textbf{Waveform}} &
\parbox[t]{0.18\textwidth}{Approximate expensive numerical-relativity waveforms across an 8D binary-black-hole parameter space.} &
\parbox[t]{0.20\textwidth}{High-dimensional time-series surrogate $h_{\rm cp}(\lambda;t)$.} &
\parbox[t]{0.15\textwidth}{Noise-weighted waveform mismatch $\mathcal{M}$.} &
\parbox[t]{0.28\textwidth}{Structured time-series modeling, waveform alignment, and stringent mismatch requirements.} \\[3pt]

\parbox[t]{0.10\textwidth}{\textbf{Remnant}} &
\parbox[t]{0.18\textwidth}{Predict the recoil velocity of the black hole produced after merger.} &
\parbox[t]{0.20\textwidth}{Regression model $\hat v_k(\lambda)$.} &
\parbox[t]{0.15\textwidth}{Normalized regression error $\mathcal{L}_{\rm rem}$.} &
\parbox[t]{0.28\textwidth}{Nonlinear dependence on mass ratio and spins, including rare high-recoil configurations.} \\[3pt]

\parbox[t]{0.10\textwidth}{\textbf{Dynamics}} &
\parbox[t]{0.18\textwidth}{Model the orbital evolution of eccentric, spinning binary black holes.} &
\parbox[t]{0.20\textwidth}{Time-series model for the PN orbital variable $x(t;\lambda)$.} &
\parbox[t]{0.15\textwidth}{Relative error $\mathcal{L}_{\rm dyn}$.} &
\parbox[t]{0.28\textwidth}{Coupled dynamical structure generated by eccentricity, spin, and orbital phase.} \\[3pt]

\parbox[t]{0.10\textwidth}{\textbf{Ringdown}} &
\parbox[t]{0.18\textwidth}{Interpolate Kerr black-hole quasi-normal-mode frequencies.} &
\parbox[t]{0.20\textwidth}{Model for $\omega(\chi_f,\ell,m,n)$.} &
\parbox[t]{0.15\textwidth}{Relative frequency error $\mathcal{L}_{\rm ring}$.} &
\parbox[t]{0.28\textwidth}{High-precision interpolation, particularly near extremal black-hole spin.} \\[3pt]

\parbox[t]{0.10\textwidth}{\textbf{Validity}} &
\parbox[t]{0.18\textwidth}{Predict where an existing waveform surrogate becomes inaccurate.} &
\parbox[t]{0.20\textwidth}{Error model $\hat{\mathcal{M}}(q,\chi_{1z},\chi_{2z},\omega_0)$.} &
\parbox[t]{0.15\textwidth}{Log-space prediction error $\mathcal{L}_{\rm val}$.} &
\parbox[t]{0.28\textwidth}{Learning an error landscape spanning several orders of magnitude rather than the physical observable itself.} \\[3pt]

\parbox[t]{0.10\textwidth}{\textbf{Analytic}} &
\parbox[t]{0.18\textwidth}{Construct an explicit analytic approximation to a binary-black-hole waveform.} &
\parbox[t]{0.20\textwidth}{Closed-form expression $h(t;q)$ using elementary functions.} &
\parbox[t]{0.15\textwidth}{Noise-weighted waveform mismatch $\mathcal{M}$.} &
\parbox[t]{0.28\textwidth}{The artifact must remain genuinely analytic; learned/SVD/PCA representations are explicitly disallowed.} \\[3pt]

\parbox[t]{0.10\textwidth}{\textbf{Template Bank}} &
\parbox[t]{0.18\textwidth}{Construct a compact set of representative waveforms that efficiently covers a search parameter space.} &
\parbox[t]{0.20\textwidth}{Ordered bank of waveform parameters $\{\lambda_i\}$.} &
\parbox[t]{0.15\textwidth}{Relative efficiency $N_{\rm domain}/N_{\rm agent}$ at fixed hidden-test coverage.} &
\parbox[t]{0.28\textwidth}{Scientific optimization balancing coverage, redundancy, and computational cost rather than standard prediction accuracy.} \\[3pt]

\parbox[t]{0.10\textwidth}{\textbf{New Physics}} &
\parbox[t]{0.18\textwidth}{Construct a $(2,2)$-mode waveform incorporating RG-resummed tail effects.} &
\parbox[t]{0.20\textwidth}{Analytic waveform construction and standalone frequency-domain implementation $h(f;\mathcal{M}_c,\eta,d_L,t_c,\phi_c,\lambda_{\rm RG})$.} &
\parbox[t]{0.15\textwidth}{Noise-weighted waveform mismatch $\mathcal{M}$.} &
\parbox[t]{0.28\textwidth}{Multi-stage formula-to-waveform reasoning requiring consistent approximations, units, conventions, and executable implementation.} \\

\bottomrule
\end{tabular}
\end{table*}

%----------------------------------------------------------------------------------------------
\subsection{Example Agent Prompt}
\label{sec:agent_prompt}
%----------------------------------------------------------------------------------------------

To make the agentic setup concrete, we show below a representative system-level prompt used to initialize one of the evaluated configurations, Opus~4.6. The prompt establishes the sequence of scientific tasks, directs the agent to the benchmark-specific instructions, and specifies the completion protocol.

\begin{center}
\fbox{%
\begin{minipage}{0.96\linewidth}
\ttfamily\small
/ralph-loop:ralph-loop "You are the Opus 4.6 agent for the gwBenchmarks suite. Your agent ID is 'opus46'. Run all eight benchmarks sequentially in this order: waveform, remnant, dynamics, ringdown, validity, analytic, template\_bank, new\_physics. For each benchmark: (1) run python llm\_agents/generate\_prompt.py opus46 <benchmark> --write from the gwBenchmarks/ root to generate your task prompt, (2) read llm\_agents/results/opus46/<bench>/AGENT\_PROMPT.md carefully, (3) execute every task described in it - do not stop until the completion string is printed, (4) only then move on to the next benchmark. Completion strings: WAVEFORM\_BENCH\_COMPLETE, REMNANT\_BENCH\_COMPLETE, DYNAMICS\_BENCH\_COMPLETE, RINGDOWN\_BENCH\_COMPLETE, VALIDITY\_BENCH\_COMPLETE, ANALYTIC\_BENCH\_COMPLETE, TEMPLATE\_BANK\_BENCH\_COMPLETE, NEW\_PHYSICS\_BENCH\_COMPLETE." --max-iterations 3
\end{minipage}%
}
\end{center}

The system-level prompt provides task sequencing and completion conditions, while each benchmark-specific prompt defines the scientific objective, available data, required outputs, constraints, and evaluation procedure. The complete benchmark-specific prompts are substantially longer and are therefore omitted from the main text; they are provided in the public repository and documented further in Appendix~\ref{sec:individual_benchmarks}. Details of the execution protocol, stopping criteria, and computational-resource differences among agent configurations are given in Appendix~\ref{app:stopping}.

%==============================================================================================
\subsection{Stopping Criteria and Resource Budgets}
\label{app:stopping}
%==============================================================================================

Long-horizon agent benchmarks must define when a task is complete and, consequently, how long an agent is allowed to continue working. In \texttt{gwBenchmarks}, stopping is determined by completion of the required scientific workflow rather than by a uniform computational budget. Each benchmark prompt specifies a \texttt{Completion Criteria} checklist requiring at least 20 distinct modeling approaches, at least three reparameterizations, coverage of all four prescribed approach categories, executed symbolic-regression runs, and a required set of output files. Once these criteria are satisfied, the agent prints a benchmark-specific completion string. We impose no common wall-clock, interaction-turn, or token budget across agent configurations.

Six configurations executed through the Claude Code CLI additionally ran inside a Ralph loop with \texttt{-{}-max-iterations 3}. This setting does not correspond to three modeling iterations or to a fixed compute budget: the loop may re-submit the prompt when an agent attempts to terminate, thereby limiting the number of continuation cycles rather than the amount of work performed within each invocation. Configurations executed through the Codex, Gemini, and OpenCode CLIs did not use this wrapper and terminated after their own invocation. The execution environment was therefore not fully uniform across configurations, and the resulting comparison should not be interpreted as compute-normalized.

The saved scorecards indicate, however, that the completion checklist generally determined stopping behavior. Among 87 agent--benchmark pairs with per-approach scorecards, 69 contain 20--24 approaches and 32 terminate exactly at the required minimum of 20, while only one contains fewer than 20. Conversely, Sonnet~4.6 explored 42--49 approaches on six benchmarks despite operating under the Ralph loop, whereas several configurations without the loop stopped at exactly 20 approaches. Approach counts therefore primarily reflect compliance with the completion protocol rather than a normalized measure of computational effort or scientific exploration. Consistent with this interpretation, the median Spearman correlation between approach index and agent-reported loss is $0.00$.

Existing agent benchmarks adopt different conventions for controlling computational effort. METR combines wall-clock limits with high token ceilings,\footnote{\url{https://metr.org/evaluations/openai-o3-report/}} SWE-Bench Pro limits interaction turns,\footnote{\url{https://scaleapi.github.io/SWE-bench_Pro-os/}} Humanity's Last Exam constrains completion length,\footnote{\url{https://github.com/centerforaisafety/hle}} and Epoch AI's FrontierMath evaluation imposes a total-token budget per run.\footnote{\url{https://epoch.ai/benchmarks/frontiermath-tiers-1-3-v2}} These choices normalize different aspects of agentic computation and therefore impose different constraints on long-horizon scientific work. Our completion-based protocol instead standardizes the required scientific workflow and deliverables while allowing resource usage to vary. Accordingly, all valid artifacts are evaluated using the same task-specific scientific criteria, but the present results should be interpreted as comparisons of artifacts produced under the specified execution protocols rather than as controlled comparisons at equal compute.

%==============================================================================================
%==============================================================================================
\section{Evaluation Framework}
\label{sec:evaluation_protocol}
%==============================================================================================
%==============================================================================================

The central design principle is to separate artifact generation from scientific verification. Agents autonomously interact with the computational environment by inspecting data, writing and executing code, constructing models, evaluating intermediate results, and revising their approaches. The final output is a scientific artifact, such as a surrogate, analytic expression, template bank, or waveform implementation. Because successful execution and agent-reported performance do not by themselves establish scientific validity, we treat the agent's own evaluations as part of the auditable work record rather than as final benchmark scores. Submitted artifacts are instead evaluated using a task-specific, pre-specified procedure, which we call the \emph{canonical evaluation}.

%----------------------------------------------------------------------------------------------
\subsection{Three Levels of Scientific-Agent Evaluation}
\label{sec:evaluation_levels}
%----------------------------------------------------------------------------------------------

We distinguish three complementary levels of scientific-agent evaluation. \emph{Integrity} asks whether the submitted artifact and accompanying work record satisfy the benchmark protocol, including task constraints, required data coverage, artifact validity, and support for reported claims. \emph{Outcome} measures the scientific performance of a valid artifact using the prescribed task-specific metric. \emph{Diagnosis} examines the saved work record for auditable evidence of failures in task understanding, method selection, implementation, evaluation, physics, verification, and reporting. These levels capture distinct aspects of agent performance: executable code may fail integrity checks, a valid artifact may perform poorly under canonical evaluation, and similar outcome scores may arise from substantially different scientific processes.

Conceptually, the framework separates three questions:
\begin{equation}
\begin{aligned}
\text{Integrity:}  &\quad \text{``Is it valid?''} \\
\text{Outcome:}    &\quad \text{``How well does it work?''} \\
\text{Diagnosis:}  &\quad \text{``What happened?''}
\end{aligned}
\end{equation}
Integrity determines whether an artifact can be meaningfully evaluated and whether its accompanying record supports the claims made about it. Outcome quantifies its scientific performance, while diagnosis characterizes evidence in the work record about how that result was produced. Section~\ref{sec:results} reports scientific outcomes and integrity failures, while Section~\ref{sec:failure_analysis} examines diagnostic evidence from the agents' work records.

%----------------------------------------------------------------------------------------------
\subsection{Independent Scientific Verification}
\label{sec:canonical_evaluation}
%----------------------------------------------------------------------------------------------

Each benchmark defines an executable canonical metric (Table~\ref{tab:benchmark_summary}), including waveform mismatch, normalized or relative prediction error, quasi-normal-mode frequency error, and template-bank efficiency. Submitted artifacts are reevaluated on the complete validation set using the same task-specific procedure across agent configurations. The verification pipeline is implemented outside the agents' writable workspace: agents may inspect the evaluation procedures but cannot modify them.

Canonical scores are determined by fixed \texttt{gwbenchmarks.metrics} implementations rather than by an LLM's assessment of scientific quality. A separate Opus~5 process orchestrates the evaluation by executing submitted artifacts against the validation data through these fixed metric implementations. Before scoring, the pipeline checks validation-set coverage, required artifact format and methodology, and the absence of placeholder or hardcoded outputs. Submissions that violate required artifact formats or methodological constraints are marked invalid rather than ranked by scientific performance. Discrepancies between an agent's reported performance and canonical evaluation are recorded as integrity failures when appropriate, while a valid and executable artifact can still be scored independently of the agent's own assessment. Agents may optimize intermediate objectives during development, but only the canonical metrics determine final scientific performance.

%----------------------------------------------------------------------------------------------
\subsection{Diagnostic Analysis of Agent Work Records}
\label{sec:diagnostic_protocol}
%----------------------------------------------------------------------------------------------

Diagnosis is performed on saved artifacts from the agents' work, including scorecards, model metadata, training and prediction code, changelogs, evaluation reports, and calibration values. We refer collectively to these materials as the \emph{work record}; they provide auditable evidence of the agent's scientific process but should not be interpreted as a complete record of its internal reasoning.

Fixed Python scripts examine these records for predefined indicators of failures in task understanding, method selection, implementation, evaluation, physics, verification, and reporting. After all agent runs were complete, Opus~5 and GPT-5.6 Sol independently produced two sets of diagnostic scripts; neither model directly judged or annotated individual work records. Once defined, the scripts are deterministic: applying the same script to the same files produces the same findings. Each finding records the failure category, supporting numerical or textual evidence, and a confidence label. The authors selected the thresholds and search patterns, so these choices remain explicit assumptions of the diagnostic analysis rather than learned judgments. Appendix~\ref{app:diagnostic_protocol} specifies the unit of analysis, treatment of multiple findings, changelog searches, and differences between the Opus~5 and GPT-5.6 Sol diagnostic pipelines.

%==============================================================================================
%==============================================================================================
\section{Results}
\label{sec:results}
%==============================================================================================
%==============================================================================================

We organize the results according to the three levels defined in Section~\ref{sec:evaluation_levels}. We first report independently measured scientific outcomes, then examine how integrity checks change their interpretation, and finally use the saved work records to diagnose failures in the scientific process.

%----------------------------------------------------------------------------------------------
\subsection{Outcome: Independently Measured Scientific Performance}
\label{sec:scientific_outcome}
%----------------------------------------------------------------------------------------------

\begin{figure*}[t]
\centering
\includegraphics[width=\linewidth]{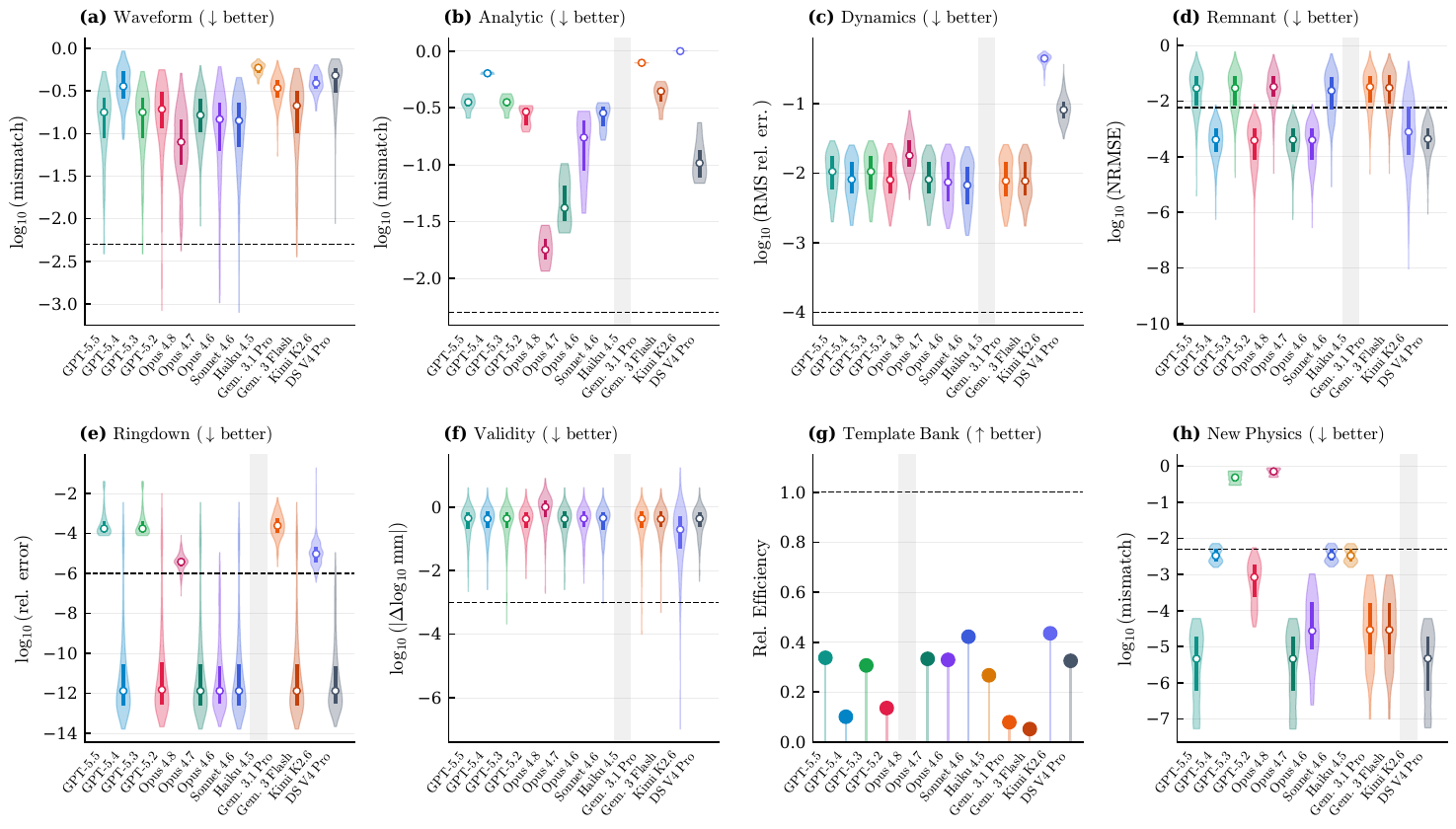}
\caption{
\textbf{Canonical evaluation reveals broad distributions of scientific performance.}
Each panel corresponds to one of the eight scientific tasks (a--h). Metrics include waveform mismatch (analogous to cosine distance between normalized waveforms), relative error, normalized RMSE, and relative efficiency. Violins represent empirical distributions over validation samples, white markers denote medians, and thick bars indicate interquartile ranges. Dashed horizontal lines indicate physics-based target thresholds where defined, and shaded vertical bands group model families for comparison. The full distributions expose behavior that aggregate scores conceal, including broad waveform errors, heavy-tailed remnant errors, near machine-precision convergence for ringdown interpolation, and large variability in tasks requiring structured modeling or physics implementation. Section~\ref{sec:scientific_outcome} discusses these scientific outcomes; Appendix~\ref{app:metrics} defines the metrics and target thresholds.
}
\label{fig:violin}
\end{figure*}

We report outcomes for 13 scored LLM coding-agent configurations spanning GPT, Claude, Gemini, and other model families, including open- and closed-weight models. Two additional configurations were attempted but excluded by the integrity checks described in Section~\ref{sec:canonical_evaluation}. Figure~\ref{fig:violin} shows the distributions of canonical per-sample scores, while Table~\ref{tab:main_results} reports their median values. Metrics are error-based, with lower values indicating better performance, except for Template Bank, where higher relative efficiency is better. Missing entries correspond to invalid or unavailable submissions.

%==============================================================================================
%==============================================================================================

Canonical evaluation reveals three broad patterns across tasks. First, performance is strongly task-dependent: no agent consistently dominates, and the tasks range from reliably solved low-dimensional interpolation to waveform and analytic modeling that remain one to two orders of magnitude from domain requirements. For Waveform, the best mismatch is $\sim30\times$ the physics target and nearly two orders of magnitude above the NR error floor. Second, full-set evaluation reveals behavior hidden by a single aggregate score. Waveform errors span nearly two orders of magnitude, and the 95th-percentile Remnant error is $\sim7\times$ its median. Third, protocol compliance changes how scores should be interpreted: the lowest-loss Analytic artifact violates the closed-form constraint, whereas the best compliant artifact uses a physics-informed hybrid expression.

The task-level results also reveal qualitatively different scientific strategies. Five agents independently converge to cubic-spline interpolation on Ringdown, producing bit-identical per-sample errors with median $\sim10^{-12}$, while GPT-5.2 uses a physics-informed coordinate transformation and also meets the stated requirement, including a near-extremal error of $3.93\times10^{-7}$. In contrast, agents construct nontrivial Analytic models (Appendix~\ref{sec:analytic_equation}) without reaching domain accuracy. On New Physics, residual errors in otherwise successful artifacts arise from numerical integration and boundary conventions, whereas larger failures trace to physics inconsistencies such as an incorrect phase prescription. Canonical scores therefore establish whether the final artifacts work, but interpreting those scores still requires integrity checks and diagnostic evidence from the work records.

%----------------------------------------------------------------------------------------------
\subsection{Integrity: Protocol and Reporting Failures}
\label{sec:eval_integrity}
%----------------------------------------------------------------------------------------------

Applying the integrity checks described in Section~\ref{sec:canonical_evaluation} reveals several cases in which agent-reported success does not constitute a valid scientific claim. We observe partial validation, metric substitution, fabricated outputs, methodological constraint violations, and execution outside the permitted benchmark scope. Table~\ref{tab:invalid_submissions} summarizes representative cases.

\begin{table*}[t]
\centering
\caption{
\textbf{Representative integrity failures identified in agent submissions and work records.}
Artifact violations determine whether a submission can be validly scored, whereas evaluation and reporting failures determine whether the available evidence supports an agent's own performance claim. Agent-reported scores are not used as final benchmark results in either case. Section~\ref{sec:eval_integrity} discusses the integrity results and their implications for interpreting scientific-agent performance; Section~\ref{sec:canonical_evaluation} describes the corresponding verification procedure.
}
\label{tab:invalid_submissions}

\footnotesize
\setlength{\tabcolsep}{5pt}
\renewcommand{\arraystretch}{1.15}

\begin{tabular}{@{}ll@{}}
\toprule
\parbox[t]{0.18\textwidth}{\textbf{Failure Type}} &
\parbox[t]{0.76\textwidth}{\textbf{Representative Example}} \\
\midrule

\parbox[t]{0.18\textwidth}{\textbf{Partial validation}} &
\parbox[t]{0.76\textwidth}{Opus 4.7 and GPT-5.4 Mini evaluated only subsets of the Waveform validation set (20 and 64 samples, respectively).} \\[3pt]

\parbox[t]{0.18\textwidth}{\textbf{Metric misuse}} &
\parbox[t]{0.76\textwidth}{Sonnet 4.6, GPT-5.5 High, and GPT-5.3 Codex used MAE-style objectives instead of required RMSE-based metrics on Remnant/Validity; multiple Waveform agents used naive FFT overlaps instead of the prescribed noise-weighted mismatch.} \\[3pt]

\parbox[t]{0.18\textwidth}{\textbf{Fabricated outputs}} &
\parbox[t]{0.76\textwidth}{Haiku 4.5 produced stub models with hardcoded losses on Dynamics, Ringdown, Validity, and Analytic; Hy3 Preview Free generated placeholder outputs rather than executable scientific models.} \\[3pt]

\parbox[t]{0.18\textwidth}{\textbf{Constraint violation}} &
\parbox[t]{0.76\textwidth}{Opus 4.8's lowest-loss Analytic artifact used a disallowed learned representation, so we excluded it from the compliant ranking. Opus 4.7 explored SVD-based candidates but ultimately submitted a compliant closed-form model.} \\[3pt]

\parbox[t]{0.18\textwidth}{\textbf{Scope violation}} &
\parbox[t]{0.76\textwidth}{Gemini 2.5 Pro modified files outside the permitted workspace, so we excluded it.} \\

\bottomrule
\end{tabular}
\end{table*}

These failures differ fundamentally from poor scientific performance and fall into two groups. Placeholder outputs, prohibited representations, and scope violations compromise the submitted artifact itself; such submissions are invalid and are omitted from the rankings. Partial validation and metric substitution instead compromise the evidence supporting an agent-reported score. If the underlying artifact remains complete and executable, canonical evaluation can still score it, but we do not accept the agent's own performance claim. This distinction matters because evaluating only a subset may omit difficult cases, while a simpler proxy metric can change the quantity being optimized. An executable artifact and a low agent-reported loss therefore do not, by themselves, establish scientific success.

More broadly, the evaluation procedure is itself part of the scientific workflow and can fail independently of model construction. Allowing an agent both to produce an artifact and to determine whether it succeeds can obscure whether apparent success reflects scientific performance or unreliable self-evaluation. Independent canonical evaluation separates these questions and provides a common basis for comparing heterogeneous scientific artifacts. In this framework, \emph{integrity} establishes whether an artifact can be validly evaluated and whether its work record supports the claims made about it, whereas \emph{outcome} measures its independently verified scientific performance. We next turn to \emph{diagnosis}, which examines evidence in the saved work records about the scientific process that produced these outcomes.

%==============================================================================================
\begin{table*}
\caption{
\textbf{Median scientific performance across the eight \texttt{gwBenchmarks} tasks.}
The table reports median per-sample validation results for the 13 scored agent configurations. Lower values indicate better performance except for Template Bank, where higher relative efficiency is better. Bold values indicate the best compliant median for each task. ``--'' denotes missing or invalid outputs; for Opus~4.8, the template bank does not reach 50\% hidden-test coverage and its relative efficiency is therefore undefined. Section~\ref{sec:scientific_outcome} discusses the task-level results and comparisons across agent configurations; Appendix~\ref{app:metrics} provides the complete metric definitions.
}
\centering
\small
\setlength{\tabcolsep}{2.5pt}
\renewcommand{\arraystretch}{1.1}
\resizebox{\linewidth}{!}{
\begin{tabular}{lcccccccc}
\toprule
Agent & Waveform & Remnant & Dynamics & Ringdown & Validity & Analytic & Template Bank & New Physics \\
\midrule
GPT-5.5 High & $1.782\times10^{-1}$ & $2.891\times10^{-2}$ & $1.053\times10^{-2}$ & $1.718\times10^{-4}$ & $4.442\times10^{-1}$ & $3.550\times10^{-1}$ & $9.030\times10^{-2}$ & $\mathbf{4.631\times10^{-6}}$ \\
GPT-5.4 Mini & $3.597\times10^{-1}$ & $4.084\times10^{-4}$ & $8.123\times10^{-3}$ & $\mathbf{1.306\times10^{-12}}$ & $4.261\times10^{-1}$ & $6.380\times10^{-1}$ & $8.411\times10^{-2}$ & $3.318\times10^{-3}$ \\
GPT-5.3 Codex & $1.782\times10^{-1}$ & $2.913\times10^{-2}$ & $1.053\times10^{-2}$ & $1.718\times10^{-4}$ & $4.350\times10^{-1}$ & $3.550\times10^{-1}$ & $3.034\times10^{-1}$ & $4.864\times10^{-1}$ \\
GPT-5.2 & $1.935\times10^{-1}$ & $\mathbf{3.850\times10^{-4}}$ & $8.022\times10^{-3}$ & $1.490\times10^{-12}$ & $4.230\times10^{-1}$ & $2.921\times10^{-1}$ & $8.133\times10^{-2}$ & $8.409\times10^{-4}$ \\
\midrule
Opus 4.8 & $\mathbf{7.973\times10^{-2}}$ & $3.221\times10^{-2}$ & $1.794\times10^{-2}$ & $3.720\times10^{-6}$ & $1.001$ & $1.780\times10^{-2}$ & -- & $7.129\times10^{-1}$ \\
Opus 4.7 & $1.648\times10^{-1}$ & $4.084\times10^{-4}$ & $8.123\times10^{-3}$ & $\mathbf{1.306\times10^{-12}}$ & $4.261\times10^{-1}$ & $\mathbf{4.161\times10^{-2}}$ & $6.444\times10^{-2}$ & $4.647\times10^{-6}$ \\
Opus 4.6 & $1.475\times10^{-1}$ & $3.950\times10^{-4}$ & $7.408\times10^{-3}$ & $1.313\times10^{-12}$ & $4.368\times10^{-1}$ & $1.735\times10^{-1}$ & $9.574\times10^{-2}$ & $2.733\times10^{-5}$ \\
Sonnet 4.6 & $1.422\times10^{-1}$ & $2.348\times10^{-2}$ & $\mathbf{6.710\times10^{-3}}$ & $\mathbf{1.306\times10^{-12}}$ & $4.493\times10^{-1}$ & $2.852\times10^{-1}$ & $2.697\times10^{-2}$ & $3.321\times10^{-3}$ \\
Haiku 4.5 & $5.931\times10^{-1}$ & -- & -- & -- & -- & -- & $2.673\times10^{-1}$ & $3.318\times10^{-3}$ \\
\midrule
Gemini 3.1 Pro & $3.425\times10^{-1}$ & $3.208\times10^{-2}$ & $7.719\times10^{-3}$ & $2.409\times10^{-4}$ & $4.327\times10^{-1}$ & $7.890\times10^{-1}$ & $9.000\times10^{-3}$ & $2.924\times10^{-5}$ \\
Gemini 3 Flash & $2.127\times10^{-1}$ & $2.988\times10^{-2}$ & $7.699\times10^{-3}$ & $\mathbf{1.306\times10^{-12}}$ & $4.195\times10^{-1}$ & $4.432\times10^{-1}$ & $4.306\times10^{-2}$ & $2.924\times10^{-5}$ \\
\midrule
Kimi K2.6 & $3.912\times10^{-1}$ & $7.796\times10^{-4}$ & $4.447\times10^{-1}$ & $9.318\times10^{-6}$ & $\mathbf{1.967\times10^{-1}}$ & $1.000$ & $\mathbf{4.286\times10^{-1}}$ & -- \\
DeepSeek V4 Pro & $4.830\times10^{-1}$ & $4.340\times10^{-4}$ & $8.126\times10^{-2}$ & $1.313\times10^{-12}$ & $4.334\times10^{-1}$ & $1.034\times10^{-1}$ & $4.655\times10^{-2}$ & $4.738\times10^{-6}$ \\
\bottomrule
\end{tabular}}
\label{tab:main_results}
\end{table*}
%==============================================================================================

%----------------------------------------------------------------------------------------------
\subsection{Diagnosis: Evidence from Agent Work Records}
\label{sec:failure_analysis}
%----------------------------------------------------------------------------------------------

\begin{figure*}[t]
\centering
\includegraphics[width=\textwidth]{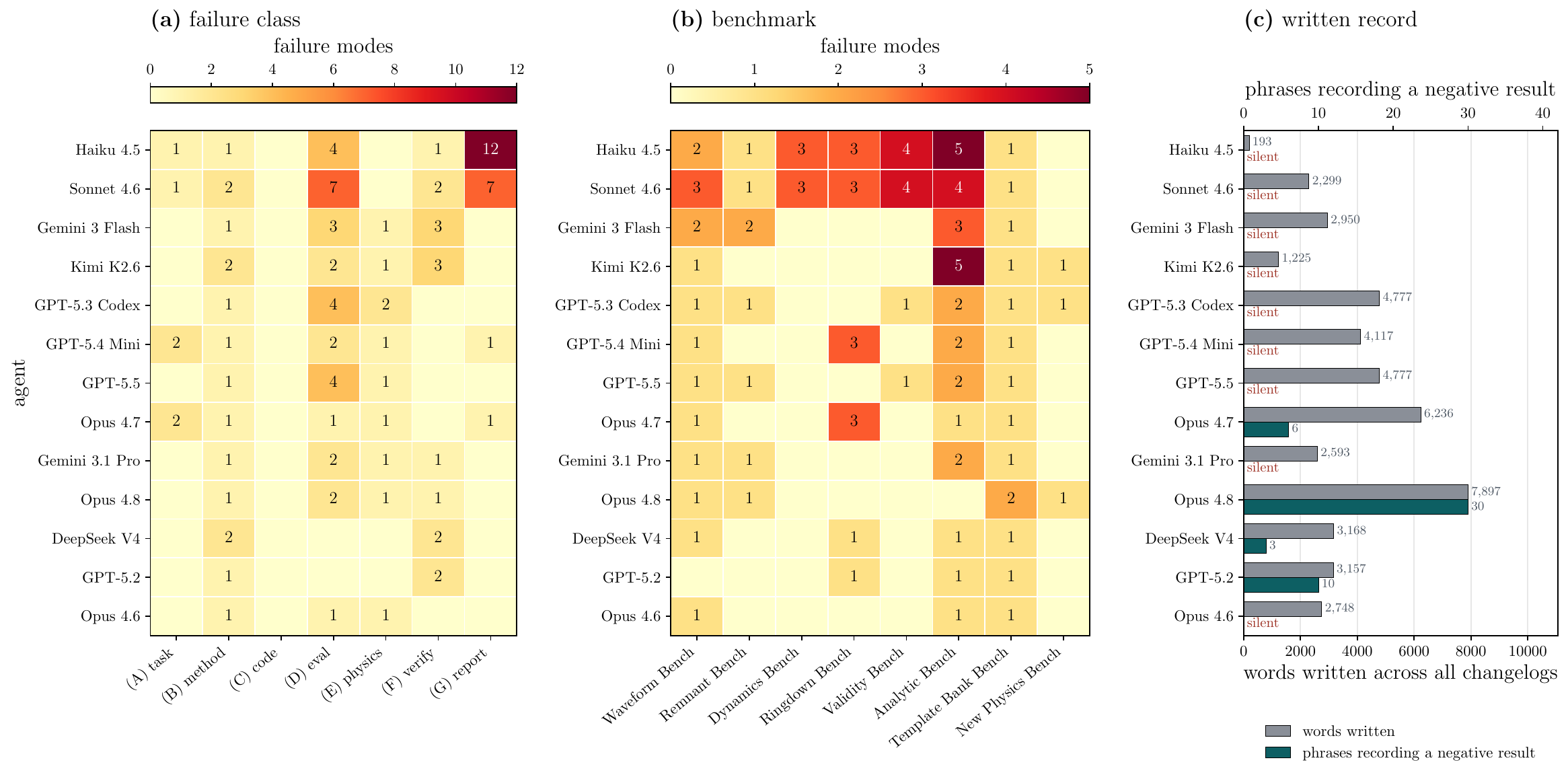}
\caption{
\textbf{Diagnostic evidence from agent work records.}
Fixed Python scripts analyze saved code, scorecards, reports, and changelogs for predefined indicators of scientific-process failures.
\textbf{(a)} Findings are grouped into seven stages of the scientific workflow: (A) task understanding, (B) method selection, (C) code, (D) evaluation, (E) physics, (F) verification, and (G) reporting.
\textbf{(b)} Distribution of findings across the eight \texttt{gwBenchmarks} tasks; each cell reports the number of findings, with darker colors indicating more findings.
\textbf{(c)} Changelog evidence comparing total word count with matches to 21 search patterns for negative results; ``silent'' denotes no match. The Analytic task contains the most failure evidence. For nine of thirteen agents, the changelogs contain no matched negative-result phrase despite, in some cases, containing thousands of words; documentation volume alone therefore does not establish scientific self-checking. Section~\ref{sec:failure_analysis} discusses these diagnostic results; Section~\ref{sec:diagnostic_protocol} describes the diagnostic methodology, with additional implementation details in Appendix~\ref{app:diagnostic_protocol}.
}
\label{fig:failure_analysis}
\end{figure*}

Figure~\ref{fig:failure_analysis}(a) groups findings from the fixed diagnostic scripts into seven stages of the scientific workflow: (A) task understanding, (B) method selection, (C) code, (D) evaluation, (E) physics, (F) verification, and (G) reporting. The evidence extends beyond code generation: saved artifacts expose problems in evaluation, verification, methodology, physics, and reporting even when a model executes successfully. Figure~\ref{fig:failure_analysis}(b) shows how these findings vary across the eight benchmarks. Different tasks stress different parts of the scientific workflow, and strong performance on one task does not imply an absence of diagnostic evidence on another. Aggregate scientific performance alone therefore does not establish reliability.

The analysis reveals an important distinction between failures of \emph{execution} and failures of \emph{scientific method}. Coding errors may prevent an agent from producing an artifact, whereas incorrect method selection, physical assumptions, or evaluation procedures can yield executable and apparently plausible outputs that are nevertheless scientifically incorrect. Verification is therefore particularly important: an apparently complete workflow can preserve a methodological or physical error in its final artifact.

%----------------------------------------------------------------------------------------------
\subsubsection{How Agent Work Records Were Analyzed}
\label{app:diagnostic_protocol}
%----------------------------------------------------------------------------------------------

The diagnostic analysis uses only released work records: \texttt{scorecard.json}, \texttt{best\_model.json}, training and prediction code, \texttt{CHANGELOG.md}, \texttt{evaluation\_report.json}, and reference values from the calibration table. It does not use interaction transcripts or attempt to infer unobserved reasoning. We analyze each agent--benchmark pair separately; 87 pairs across the 15 attempted agent configurations contain per-approach scorecards. A single pair may produce multiple findings, including multiple findings within the same category. Each finding records a specific problem type, the numerical or textual evidence supporting it, a confidence level (certain, strong, or indicative), and \texttt{n\_approaches} when a finding applies to multiple approaches.

Two diagnostic implementations were independently developed after completion of the agent runs. The Opus~5 implementation is \path{additional/taxonomy/classify_failures.py}, while the GPT-5.6 Sol implementation is \path{additional_codex/04_failure_taxonomy/scripts/classify_artifact_failures.py}. Both use fixed, evidence-based rules rather than model judgments at runtime. The Opus~5 script identifies, for example, partial validation from \texttt{n\_val}, missing required loss components, stub prediction files, arithmetic patterns in reported losses, and discrepancies between claimed values and those obtained by canonical evaluation. The GPT-5.6 Sol implementation performs the same general class of deterministic analysis but searches for a different set of predefined problems.

Neither diagnostic implementation calls an LLM or human judge at runtime. The Opus~5 analysis produces 104 findings across all released work records, of which the 100 findings associated with the 13 configurations retained for outcome comparison are shown in Figure~\ref{fig:failure_analysis}. The independently developed GPT-5.6 Sol analysis produces 130 findings. Because the two implementations search for different sets of problems, their category counts can differ substantially---for example, they identify 34 versus 78 evaluation findings and 10 versus 1 physical-validity findings. We therefore do not treat them as independent annotations of a common labeling task and do not report an inter-rater agreement statistic. A controlled comparison in which both implementations apply identical failure definitions to the same agent--benchmark pairs remains future work.

For the changelog analysis in Figure~\ref{fig:failure_analysis}(c), a separate fixed script applies 21 case-insensitive regular-expression patterns to each \texttt{CHANGELOG.md} and sets \texttt{acknowledges} when at least one pattern matches. Re-running any fixed script on the same released files reproduces its output exactly. The resulting findings are deterministic conditional on the specified rules, but the thresholds and search patterns themselves were selected by the authors and should therefore be regarded as assumptions of the diagnostic analysis.

%----------------------------------------------------------------------------------------------
\subsection{Physics-Informed Novelty in Agent-Constructed Models}
\label{sec:physics_informed_modeling}
%----------------------------------------------------------------------------------------------

Final benchmark scores measure whether an artifact works, but they do not fully characterize \emph{how} an agent solves a scientific problem. Inspection of the submitted models reveals several cases in which agents introduce scientifically meaningful structure rather than relying only on generic interpolation, regression, or machine-learning techniques. We refer to this as \emph{physics-informed novelty}: a modeling choice generated by the agent that exploits recognizable structure of the underlying physical problem and is not explicitly prescribed by the benchmark.

We emphasize that ``novelty'' here does not imply the discovery of new physical laws. Rather, it refers to nontrivial scientific choices made during model construction. We highlight two representative examples. In the Analytic Bench, Opus~4.7 constructs a single closed-form inspiral--merger--ringdown model by combining physically motivated descriptions of different stages of binary-black-hole evolution. In the Ringdown Bench, GPT-5.2 independently introduces a nonlinear transformation of the black-hole spin that smooths the difficult near-extremal region of the interpolation problem. The first example concerns \emph{model architecture}; the second concerns \emph{physical representation of the input space}. Together, they show that agent behavior can contain scientifically interesting structure that is not captured by aggregate benchmark scores alone.

%----------------------------------------------------------------------------------------------
\subsubsection{A Physics-Informed Analytic Inspiral--Merger--Ringdown Model}
\label{sec:analytic_equation}
%----------------------------------------------------------------------------------------------

The Analytic Bench asks agents to construct an explicit mathematical representation of the dominant gravitational waveform produced by a binary-black-hole merger. This is deliberately more restrictive than conventional surrogate modeling: learned representations such as reduced bases, singular-value decompositions (SVDs), or principal-component decompositions (PCAs) are prohibited. The final model must instead remain an explicit analytic function of time and binary parameters.

Opus~4.7 constructs such a model by separating the waveform into amplitude and phase,
\begin{equation}
h_{22}(t;q) = \exp\!\bigl[\log A(t;q)\bigr]\,\exp\!\bigl[-i\phi(t;q)\bigr],
\end{equation}
where $q$ is the binary mass ratio. The agent introduces the transformed variables
\begin{equation}
x = \ln q, \qquad \tau = \frac{2(t+2500)}{2575}-1,
\end{equation}
which map the parameter and time dependence into coordinates convenient for analytic approximation.

Rather than fitting the entire waveform with a single generic function, the model explicitly distinguishes the pre- and post-merger regimes. These regimes are connected through smooth transition functions,
\begin{equation}
\sigma_+(t) = \frac{1}{2}\left[1+\tanh(t/5)\right], \qquad \sigma_-(t) = 1-\sigma_+(t),
\end{equation}
with analogous phase-dependent functions
\begin{equation}
\sigma_+^{\phi}(t;q) = \frac{1}{2}\left[1+\tanh\!\left(\frac{t}{T_b(q)}\right)\right], \qquad \sigma_-^{\phi}(t;q) = 1-\sigma_+^{\phi}(t;q).
\end{equation}

The resulting amplitude and phase consist of physically structured baseline models supplemented by flexible analytic corrections,
\begin{align}
\log A(t;q) &= \log A_{\rm base}(t;q) + \sum_{n=0}^{13}a_n(q)T_n(\tau), \\
\phi(t;q) &= \phi_{\rm base}(t;q) + \sum_{n=0}^{13}b_n(q)T_n(\tau),
\end{align}
where $T_n(\tau)$ denotes the $n$th Chebyshev polynomial. From a machine-learning perspective, the baseline provides an explicit inductive bias about the structure of the waveform, while the Chebyshev terms provide additional approximation capacity.

The baseline amplitude connects a slowly evolving pre-merger signal to a decaying post-merger signal,
\begin{equation}
\begin{aligned}
\log A_{\rm base}(t;q) &= \log A_{\rm pk}(q) - \sigma_-\,p_{\rm pre}(q)\ln\!\left(1-\frac{t}{T_{\rm pn}(q)}\right) \\
&\quad - \sigma_+\left(\frac{\max(t,0)}{\tau_{\rm RD}(q)}\right)^{p_{\rm post}(q)},
\end{aligned}
\end{equation}
while the baseline phase transitions from an inspiral-like chirping evolution to a late-time oscillation characterized by a ringdown frequency $\omega_{\rm RD}$,
\begin{equation}
\begin{aligned}
\phi_{\rm base}(t;q) &= \phi_{\rm pk}(q) - \sigma_-^{\phi}\frac{8}{5}\omega_{\rm pn}(q)T_{\rm pn}^{\phi}(q) \\
&\quad\times\left[\left(1-\frac{t}{T_{\rm pn}^{\phi}(q)}\right)^{5/8}-1\right] + \sigma_+^{\phi}\omega_{\rm RD}(q)t.
\end{aligned}
\end{equation}

All scalar functions controlling this model---including quantities such as $\log A_{\rm pk}(q)$ and $\omega_{\rm RD}(q)$ and the Chebyshev coefficients $a_n(q)$ and $b_n(q)$---are represented as fourth-degree polynomials in $x=\ln q$,
\begin{equation}
c(q) = \sum_{k=0}^{4}c_k x^k.
\end{equation}
The complete construction contains 38 scalar functions of $q$, each represented by five coefficients, for a total of 190 scalar coefficients. This parameter count is not especially small by conventional regression standards. The important property is instead \emph{representational}: the submitted waveform remains a single explicit function composed of elementary functions and Chebyshev polynomials, satisfying the Analytic Bench requirement without using a learned latent representation.

\begin{figure}[t]
\centering
\includegraphics[width=\linewidth]{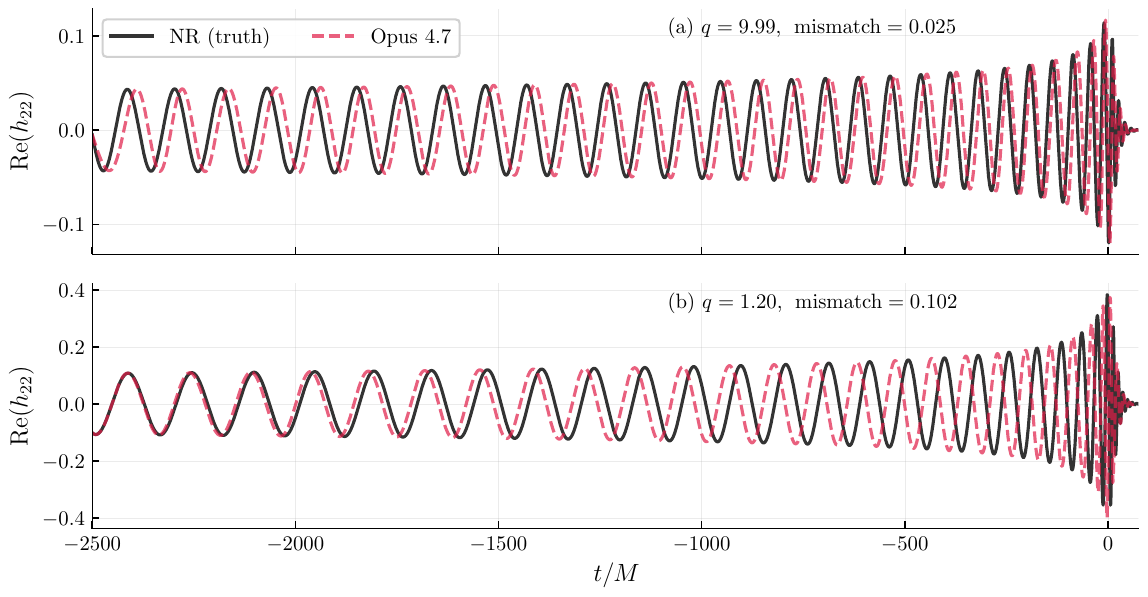}
\caption{
\textbf{Physics-informed analytic waveform constructed by Opus~4.7.}
The real part of the dominant $(2,2)$ gravitational-wave mode is compared with numerical-relativity (NR) reference waveforms for two representative mass ratios: \textbf{(a)} $q=9.99$, with mismatch $\mathcal{M}=0.025$, and \textbf{(b)} $q=1.20$, with mismatch $\mathcal{M}=0.102$. The explicit analytic model reproduces the overall inspiral--merger--ringdown morphology while retaining visible differences from the NR target. Section~\ref{sec:analytic_equation} describes the model architecture and its physics-informed construction; Appendix~\ref{app:metrics} defines the waveform mismatch used for evaluation.
}
\label{fig:analytic_opus47}
\end{figure}

Figure~\ref{fig:analytic_opus47} shows that this construction reproduces the qualitative inspiral--merger--ringdown morphology across substantially different mass ratios. Its mismatches nevertheless remain well above those achievable with high-accuracy data-driven waveform surrogates. The scientific interest of this example therefore does not lie in state-of-the-art accuracy. Rather, the agent autonomously constructs a compliant model architecture that incorporates recognizable features of the physical problem: distinct inspiral and ringdown behavior, a smooth transition between regimes, and a flexible analytic correction to the physically motivated baseline.

For an ML reader, this can be viewed as an example of an agent choosing a useful \emph{inductive bias} instead of applying a generic approximator directly to the data. For a GW reader, it corresponds to decomposing the waveform according to known stages of binary evolution while retaining sufficient flexibility to fit numerical-relativity data. The resulting artifact illustrates a capability that a scalar benchmark score alone would miss: an agent can construct a scientifically interpretable approximation even when the resulting model does not yet reach precision-waveform requirements.

%----------------------------------------------------------------------------------------------
\subsubsection{A Physics-Informed Coordinate for Kerr Ringdown}
\label{sec:gpt52_coordinate}
%----------------------------------------------------------------------------------------------

A second example appears in the Ringdown Bench, which asks agents to model the quasi-normal-mode (QNM) frequencies of a rotating Kerr black hole as functions of its final dimensionless spin $\chi_f$. QNMs determine the characteristic oscillations of the remnant black hole after merger and vary particularly rapidly as the spin approaches its extremal value, $\chi_f\rightarrow1$.

Most of the highest-performing agents treated this primarily as an interpolation problem. Opus~4.6, Opus~4.7, Sonnet~4.6, GPT-5.4 Mini, and Gemini~3 Flash independently converged on cubic-spline interpolation directly in $\chi_f$, reaching median relative errors of approximately $1.3\times10^{-12}$, far below the benchmark requirement of $10^{-6}$. Their implementations differ in detail but correspond to essentially the same numerical strategy.

\begin{figure}[t]
\centering
\includegraphics[width=0.9\linewidth]{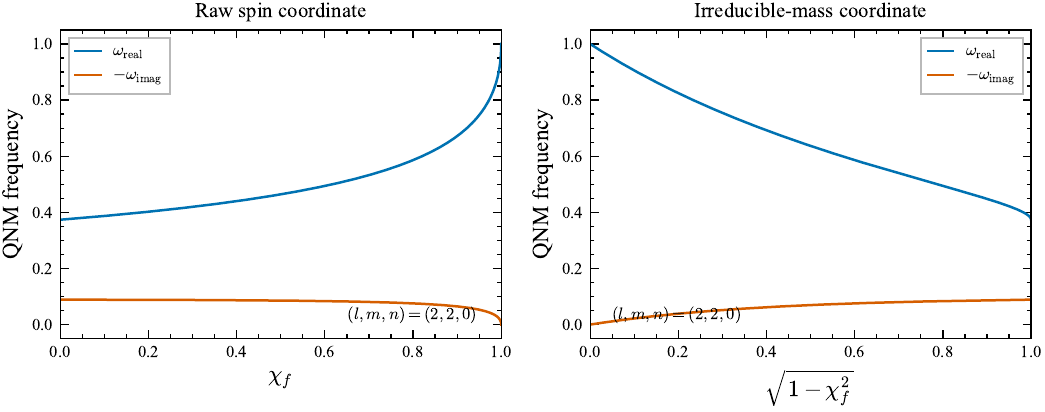}
\caption{
\textbf{Physics-informed reparameterization of the Kerr Ringdown problem.}
\textbf{Left:} The $(\ell,m,n)=(2,2,0)$ QNM frequency as a function of the usual dimensionless final spin $\chi_f$. Its increasingly steep variation as $\chi_f\rightarrow1$ makes the near-extremal region numerically challenging. \textbf{Right:} The same behavior expressed using $x_{\rm Kerr}=\sqrt{1-\chi_f^2}$, the coordinate introduced by GPT-5.2, which substantially smooths the near-extremal dependence. Section~\ref{sec:gpt52_coordinate} discusses the agent's representation choice and its connection to Kerr geometry.
}
\label{fig:qnm_coordinate_transform}
\end{figure}

GPT-5.2 instead changes the representation of the problem. Rather than interpolating directly in $\chi_f$, it introduces
\begin{equation}
x_{\rm Kerr} = \sqrt{1-\chi_f^2}.
\end{equation}
As shown in Fig.~\ref{fig:qnm_coordinate_transform}, the QNM frequency varies much more smoothly in this coordinate near the extremal-spin boundary. In ML terminology, the agent performs a physically motivated \emph{feature transformation}: it changes the input representation so that the target function becomes easier to approximate. In numerical-analysis terminology, it chooses a coordinate better adapted to the structure of the function being interpolated.

The transformation is not merely an arbitrary rescaling of the input variable. The same combination of spin appears naturally in Kerr geometry. For example, the irreducible mass of a Kerr black hole satisfies
\begin{equation}
\frac{M_{\rm irr}}{M} = \sqrt{\frac{1+\sqrt{1-\chi_f^2}}{2}},
\end{equation}
so that
\begin{equation}
x_{\rm Kerr}=2\left(\frac{M_{\rm irr}}{M}\right)^2-1.
\end{equation}
Related combinations also appear in the horizon and near-extremal properties of Kerr black holes, where QNM spectra develop characteristic behavior as $\chi_f\rightarrow1$~\cite{Yang:2012pj,Yang:2013uba}.

This representation does not achieve the smallest Ringdown error in the benchmark: direct cubic splines already interpolate the supplied data to nearly machine precision. Its significance is instead qualitative. GPT-5.2 independently selects a coordinate that both reflects the underlying black-hole geometry and regularizes a numerically difficult region of the function being modeled. The benchmark did not instruct the agent to use Kerr-motivated coordinates or to search for a near-extremal reparameterization.

We therefore do not interpret the result as the discovery of new physics. The transformation and related near-extremal variables are known in black-hole physics. The novelty is at the level of the \emph{agent's solution}: while solving an interpolation benchmark, the agent independently recovers a physically meaningful representation that is well adapted to the problem.

Taken together, the Analytic and Ringdown examples reveal a dimension of scientific-agent capability that is largely invisible to leaderboard-style evaluation. In one case, an agent constructs a waveform model around physically distinct dynamical regimes; in the other, an agent changes coordinates to expose a smoother representation of a difficult physical limit. Neither example establishes that current agents can autonomously discover new physical laws. They do, however, show that agents can sometimes introduce useful physical structure while constructing quantitative scientific models. Evaluating such behavior alongside accuracy, integrity, and failure evidence provides a richer picture of scientific capability than final benchmark scores alone.

%==============================================================================================
%==============================================================================================
\section{Discussion and Conclusion}
\label{sec:conclusion}
%==============================================================================================
%==============================================================================================

We introduced \texttt{gwBenchmarks}, an end-to-end benchmark for scientific agents in gravitational-wave physics, and used it to study both what current agents can accomplish in a quantitatively demanding scientific domain and how their outputs should be evaluated. The eight tasks span waveform approximation, remnant prediction, orbital dynamics, Kerr ringdown, waveform-validity estimation, analytic waveform construction, template-bank optimization, and implementation of new theoretical corrections (Table~\ref{tab:benchmark_summary}). They require agents to construct heterogeneous scientific artifacts---from regression models and time-series surrogates to explicit analytic expressions and executable physics implementations---that can be tested against precise domain-specific criteria. The benchmark therefore probes scientific capability beyond task completion or code execution alone, in a regime characterized by expensive reference calculations and stringent accuracy requirements (Figure~\ref{fig:verification_regime}).

The results reveal strongly task-dependent capabilities. Current agents solve some well-structured interpolation and regression problems to high accuracy, including near machine-precision interpolation of Kerr quasi-normal-mode frequencies, while demanding waveform and analytic-modeling tasks remain far from precision requirements (Figure~\ref{fig:violin} and Table~\ref{tab:main_results}). No single agent configuration dominates across all eight tasks. Final scores also miss scientifically interesting behavior. Section~\ref{sec:physics_informed_modeling} highlights examples of physics-informed novelty: Opus~4.7 constructs an explicit inspiral--merger--ringdown waveform around physically distinct dynamical regimes, while GPT-5.2 independently introduces a Kerr-motivated coordinate that smooths the near-extremal Ringdown problem. These are not discoveries of new physics, but they show that scientific agents can sometimes select useful representations and inductive biases that reflect the structure of the underlying problem.

At the same time, \texttt{gwBenchmarks} demonstrates why scientific capability cannot be assessed reliably from task completion or agent-reported performance alone. Partial validation, metric substitution, constraint violations, scope violations, and fabricated outputs produce cases in which apparent success does not support the corresponding scientific claim (Table~\ref{tab:invalid_submissions}). Conversely, an independently measured score establishes the performance of a final artifact but does not reveal problems in task understanding, methodology, implementation, evaluation, physics, verification, or reporting. We therefore separate evaluation into three complementary questions: \emph{integrity} asks whether the artifact and supporting record are valid, \emph{outcome} asks how well the artifact performs under the prescribed scientific metric, and \emph{diagnosis} examines auditable evidence about the process that produced it. Fixed task-specific metrics and diagnostic checks make these distinctions testable without requiring an LLM or human judge to determine scientific correctness at evaluation time.

These findings have implications beyond gravitational-wave modeling. In scientific workflows, executable code is often only an intermediate product: the resulting artifact must satisfy physical constraints, use the intended metric, cover the required domain, and reach an accuracy appropriate for scientific use. Improvements on general coding or reasoning benchmarks therefore need not translate directly into scientific reliability. Progress may require both stronger general agent capabilities and better integration of domain knowledge through scientific skills, literature retrieval, physically structured representations, and explicit verification. The physics-informed solutions observed here show that agents can make productive use of such structure, while the integrity and diagnostic results demonstrate why their scientific claims must remain independently testable.

Several limitations qualify these conclusions. The comparison is not compute-normalized: configurations stopped according to the prescribed completion protocol, while differences in execution wrappers and software environments resulted in unequal computational resources (Appendix~\ref{app:stopping}). The results therefore compare artifacts produced under the specified execution protocols using common scientific criteria, rather than models given equal compute. The diagnostic analysis also operates on released work records rather than complete interaction transcripts. Figure~\ref{fig:failure_analysis} reports deterministic findings from fixed scripts; the independently developed Opus~5 and GPT-5.6 Sol implementations search for different predefined problems and should not be interpreted as independent annotations or used to infer failure rates from their relative counts. Finally, gravitational-wave modeling is particularly amenable to this framework because many outputs admit precise executable metrics and high-fidelity numerical reference data; extending it to domains with less directly verifiable outcomes will require additional validation strategies.

Future versions of \texttt{gwBenchmarks} could incorporate signal detection, Bayesian parameter estimation, population inference, uncertainty quantification, and more open-ended scientific tasks. Complete interaction records could enable deeper analysis of how hypotheses, failed approaches, verification, and revision evolve during agentic research. Controlled experiments could also test whether domain-specific skills, literature access, scientific tools, or structured verification improve both scientific accuracy and reliability. Such extensions would move evaluation from well-defined modeling tasks toward increasingly autonomous scientific research.

We release the benchmark data, full per-sample results, agent-produced artifacts, and evaluation resources. The datasets are hosted on HuggingFace$^{\ref{data_link}}$, with the evaluation pipeline and released agent outputs available through GitHub$^{\ref{code_link}}$. By combining challenging end-to-end scientific tasks with artifact-level verification, \texttt{gwBenchmarks} provides both a quantitative measure of scientific-agent capability and a framework for determining when apparent success constitutes a valid scientific result.

%==============================================================================================
\begin{acknowledgments}
We thank Matias Zaldarriaga, Tejaswi Venumadhav, Scott Field, and Gaurav Khanna for useful discussions. Computational resources were provided by the Anvil supercomputer through NAIRR allocation NAIRR260298 and ACCESS allocation PHY260280.
\end{acknowledgments}
%==============================================================================================

%==============================================================================================
\bibliography{references}
%==============================================================================================

%==============================================================================================
\newpage
\appendix

\section{Gravitational-Wave Glossary}
\label{sec:glossary}

\begin{itemize}
\setlength{\itemsep}{0pt}

\item \textbf{General relativity (GR):} Einstein's theory of gravity, describing gravity as the dynamics of spacetime.

\item \textbf{Gravitational-wave (GW) astronomy:} The study of astrophysical systems through their emitted gravitational waves.

\item \textbf{Beyond-GR theories:} Extensions or alternatives to GR that can predict observable deviations in GW signals.

\item \textbf{Binary black hole (BBH):} A pair of black holes that orbit, inspiral, and eventually merge.

\item \textbf{Waveform:} The GW signal $h(t)$, typically represented as a time series with amplitude and phase.

\item \textbf{Ringdown:} The post-merger stage in which the remnant black hole settles through damped oscillations.

\item \textbf{Quasi-normal modes (QNMs):} Characteristic damped oscillations of a perturbed black hole, determined by its properties.

\item \textbf{Numerical relativity (NR):} Numerical solution of Einstein's equations, used to simulate strongly relativistic systems such as merging black holes.

\item \textbf{Post-Newtonian (PN) approximation:} An analytic expansion of GR applicable primarily to the slow-motion, weak-field inspiral.

\item \textbf{Effective-one-body (EOB) models:} Models that map binary dynamics onto an effective one-body problem and combine analytic and numerical information.

\item \textbf{GW detection:} Identification of GW signals in detector data, commonly using matched filtering.

\item \textbf{Parameter estimation:} Inference of source properties, such as masses and spins, from an observed GW signal.

\item \textbf{Mismatch:} A noise-weighted measure of disagreement between two waveforms, with zero indicating perfect agreement.

\item \textbf{Recoil (kick) velocity:} The velocity acquired by a merger remnant because GWs carry away net linear momentum.

\end{itemize}

%==============================================================================================
%==============================================================================================
\section{Details of Individual Benchmarks}
\label{sec:individual_benchmarks}
%==============================================================================================
%==============================================================================================

%==============================================================================================
\subsubsection{Waveform Bench}
\label{sec:waveform_bench}
Given the need for fast yet accurate approximations to NR waveforms, we consider the task of learning a surrogate model: a cheap emulator of expensive simulations~\citep{blackman2017precessingsurrogate,varma2019precessingsurrogate}. The target signals come from precessing binary black holes, where the black-hole spins cause the orbital plane to wobble and create amplitude and phase modulations. To make the problem tractable while retaining this structure, we represent the signal in a rotating coprecessing frame and model the dominant waveform component.
We use simulations from the public Simulating eXtreme Spacetimes (\texttt{SXS}) catalog (v3.0.0)~\citep{boyle2019sxscatalog,Scheel:2025jct}\footnote{\href{https://data.black-holes.org/simulations/index.html}{https://data.black-holes.org/simulations/index.html}}, generated with the Spectral Einstein Code (\texttt{SpEC}) and accessed via the \texttt{sxs}\footnote{\href{https://sxs.readthedocs.io/en/main/}{https://sxs.readthedocs.io/en/main/}} package. The dataset consists of 250 training and 200 validation simulations, covering mass ratios $q = m_1/m_2 \in [1,8]$ (with $m_1 \geq m_2$), spin magnitudes $|\boldsymbol{\chi}_{1,2}| \leq 0.8$, and low eccentricity ($e < 10^{-3}$), where $\boldsymbol{\chi}_i = (\chi_{ix}, \chi_{iy}, \chi_{iz})$ are the dimensionless spin vectors. The training and validation sets are chosen by a greedy coverage strategy so that both splits span a diverse range of binary configurations.
The input is an 8D parameter vector $\lambda = \{q, \boldsymbol{\chi}_1, \boldsymbol{\chi}_2, \omega_0\}$, where $\omega_0$ is a reference orbital frequency. The output is the coprecessing-frame waveform $h_{\mathrm{cp}}(\lambda; t)$ on a standardized grid $\{t_i\}$.
Agents may use interpolation, ML, or hand-designed modeling approaches. They are given practical guidance such as aligning waveforms before fitting and modeling amplitude and phase separately. Accuracy is measured by frequency-domain mismatch $\mathcal{M}$ (Appendix~\ref{app:metrics}; Eq.~\eqref{eq:fd_mismatch}), a standard GW distance between two signals, with an NR error floor of $\sim 8.4 \times 10^{-4}$.
This is one of the most challenging tasks in the benchmark suite, particularly among data-driven modeling problems, combining a high-dimensional parameter space with structured time-series outputs and strict alignment requirements.

%==============================================================================================
\subsubsection{Analytic Bench}
\label{sec:analytic_bench}
This benchmark asks whether agents can write an explicit formula for a waveform rather than fitting a black-box model. Numerical surrogates are widely used, but an analytic expression is more restrictive: it must be directly evaluable and cannot hide the solution in stored data or a learned representation.
We restrict to quasi-circular non-spinning BBH systems using \texttt{SXS} waveforms with $q \in [1,20]$, negligible spins ($|\boldsymbol{\chi}_i| < 0.01$), and low eccentricity ($e < 0.005$). The dataset contains 21 training and 20 validation simulations, where the coprecessing frame coincides with the inertial frame.
The input is $\lambda = \{q\}$, and the output is the waveform $h(t; \lambda)$ on a standardized grid $\{t_i\}$. The model must be expressed in closed form using elementary functions, with data-driven representations disallowed.
Model accuracy is evaluated using frequency-domain mismatch $\mathcal{M}$ (which is an analogue of cosine distance between unit waveform vectors, see Appendix~\ref{app:metrics}; Eq.~\eqref{eq:fd_mismatch}).
This task is challenging because the waveform changes character over time. A good formula must combine approximations for the early inspiral and the final ringdown into one coherent model.

%==============================================================================================
\subsubsection{Dynamics Bench}
\label{sec:dynamic_bench}
This benchmark evaluates whether agents can model orbital motion when the binary is not a simple circular system. Eccentricity and spin introduce extra time-dependent structure, similar to learning a dynamical system with several coupled control parameters.
Instead of predicting the full waveform, agents predict the PN frequency parameter $x(t)$, a compact time series that tracks the binary's orbital evolution. Because NR data are limited in this regime, we generate simulations using the EOB framework via the \texttt{SEOBNRv5EHM} model~\citep{gamboa2024seobnrv5ehm,paul2024eccentricspinning} from the \texttt{pyseobnr}\footnote{\href{https://waveforms.docs.ligo.org/software/pyseobnr/}{https://waveforms.docs.ligo.org/software/pyseobnr/}} package~\citep{Mihaylov:2023bkc}.
The input is $\lambda = \{q, \chi_{1z}, \chi_{2z}, e_0, \zeta_0, \omega_0\}$ with $q \in [1,6]$, $\chi_{1z}, \chi_{2z} \in [-0.6, 0.6]$, $e_0 \in [0.001, 0.5]$, $\zeta_0 \in [0,\pi]$, and $\omega_0 \in [0.0075, 0.0085]$. 
Here, $e_0$ controls how non-circular the initial orbit is, and $\zeta_0$ sets the initial orbital phase variable used for eccentric motion. 
The output is $x(t; \lambda)$ on a grid $\{t_i\}$. The dataset contains 250 simulations per split from Latin hypercube sampling. Model accuracy is evaluated using relative error $\mathcal{L}_{\mathrm{dyn}}$ (Appendix~\ref{app:metrics}; Eq.~\eqref{eq:dyn_err}).
This is a moderately challenging task, where eccentricity introduces nontrivial temporal structure.

%==============================================================================================
\subsubsection{Remnant Bench}
\label{sec:remnant_bench}
This benchmark evaluates prediction of final merger properties. Unlike the time-series tasks, this is a tabular regression problem: map the initial binary parameters to the recoil velocity $v_k$ of the final black hole.
The input uses the same parameterization $\lambda$ as in the \textbf{Waveform Bench}, and the output is $v_k(\lambda)$. The mapping is nonlinear because unequal masses and spins can cause gravitational radiation to be emitted more strongly in one direction, pushing the final black hole in the opposite direction. Kick velocities range from near zero to $\sim 3000\,\mathrm{km/s}$, making the task moderately challenging. Accurate recoil models are important for astrophysical applications such as predicting whether merger remnants remain bound in star clusters~\citep{Islam:2025drw,Islam:2026yxx}.
From an initial pool of $\sim 3855$ simulations, we construct the dataset using a greedy, coverage-optimized selection strategy, partitioning simulations into non-spinning, aligned-spin, and precessing categories to ensure uniform coverage.
Model accuracy is evaluated using normalized regression error $\mathcal{L}_{\mathrm{rem}}$ (Appendix~\ref{app:metrics}; Eq.~\eqref{eq:rem_err}).

%==============================================================================================
\subsubsection{Ringdown Bench}
\label{sec:ringdown_bench}
This benchmark is a comparatively clean interpolation task. After a merger, the final black hole settles down by emitting damped oscillations called ringdown modes. Their frequencies are smooth functions of the final black-hole spin and discrete mode labels.
The agent predicts quasi-normal mode (QNM) frequencies $(\omega_r, \omega_i)$ of Kerr black holes as functions of the final spin $\chi_f$ and mode indices $(\ell, m, n)$~\citep{berti2005spectroscopy}. Each mode defines a smooth mapping from spin to complex frequency, making this a structured regression problem.
We use high-precision QNM data from Ref.~\citep{Cook:2014cta}\footnote{\href{https://zenodo.org/records/2650358}{https://zenodo.org/records/2650358}}, which tabulates Kerr quasi-normal modes for gravitational perturbations ($s=-2$). The dataset spans $\ell \in [2,16]$, $|m| \leq \ell$, and overtones $n \in [0,7]$. For each mode, frequencies are provided over a dense grid of dimensionless spin values $\chi_f \in [0, 0.999999999]$ with approximately $1062$ samples per mode. We construct a consolidated dataset of $\sim 2280$ modes and split it into training and validation sets.
The input is $\lambda = \{\chi_f, \ell, m, n\}$, and the output is $\omega(\lambda)$. Model accuracy is evaluated using relative error $\mathcal{L}_{\mathrm{ring}}$ (Appendix~\ref{app:metrics}; Eq.~\eqref{eq:ring_err}).
This is the simplest task in the benchmark.

%==============================================================================================
\subsubsection{Validity Bench}
\label{sec:validity_bench}
This benchmark asks agents to predict when another model will fail. Instead of predicting a physical quantity directly, the agent learns an error landscape: the mismatch $\hat{\mathcal{M}}$ (which is an analogue of cosine distance) between expensive NR waveforms and a faster surrogate model (\texttt{NRHybSur3dq8}~\citep{varma2019precessingsurrogate} from the \texttt{gwsurrogate}\footnote{\href{https://github.com/sxs-collaboration/gwsurrogate/}{https://github.com/sxs-collaboration/gwsurrogate/}} package~\citep{Field:2025isp}). This kind of reliability modeling is important whenever a scientific workflow depends on approximate models.
We construct the dataset using aligned-spin, quasi-circular BBH simulations from the \texttt{SXS} catalog paired with surrogate-generated waveforms. The parameter space includes mass ratios $q \in [1,10]$ and aligned spins $(\chi_{1z}, \chi_{2z})$ with negligible in-plane components ($|\chi_{ix,y}| < 0.01$) and low eccentricity ($e < 0.01$). For each simulation, we compute the mismatch after aligning NR and surrogate waveforms in time and phase and interpolating onto a common grid. From an initial pool of $\sim 810$ candidates, we obtain 786 valid simulations, split evenly into 393 training and 393 validation samples.
The input consists of $(q, \chi_{1z}, \chi_{2z}, \omega_0)$, and the output is $\hat{\mathcal{M}}(\lambda)$. Model accuracy is evaluated using log-space error $\mathcal{L}_{\mathrm{val}}$ (Appendix~\ref{app:metrics}; Eq.~\eqref{eq:val_err}).
This is a moderately challenging task. The target errors span several orders of magnitude, so the model must capture both broad trends and localized regions where the surrogate is unusually inaccurate.

%==============================================================================================
\subsubsection{New Physics Implementation Bench}
\label{sec:new_physics_bench}

The New Physics Bench asks agents to construct a waveform that has not previously been derived in this combined form, using recently developed analytic ingredients for radiative tail effects in general relativity~\cite{Cipriani:2026xmx,Ivanov:2025ozg,Fucito:2024wlg,Goldberger:2009qd}. Unlike the regression and interpolation tasks, no target waveform data or existing implementation is provided. Instead, the agent receives a compact source packet of analytic results and must combine them into a consistent, executable waveform model.

The dominant non-spinning $(2,2)$ mode is constructed from
\begin{equation}
\widehat{h}_{22}=H_{\mathrm{eff}}\,T_{22}\,\rho_{22}^{\,2}e^{i\delta_{22}},
\end{equation}
where $H_{\mathrm{eff}}$ describes the orbital dynamics, $T_{22}$ contains the radiative tail contribution, and $\rho_{22}$ and $\delta_{22}$ provide additional amplitude and phase corrections. The complete expressions are supplied to the agents. A parameter $\lambda_{\mathrm{RG}}$ controls the RG-resummed contribution within $T_{22}$, with $\lambda_{\mathrm{RG}}=1$ corresponding to the full GR result.

Agents must then derive the orbital evolution from energy balance. Defining
\begin{equation}
M=\frac{\mathcal{M}_c}{\eta^{3/5}}, \qquad x=(\pi M f)^{2/3},
\end{equation}
the supplied $(2,2)$-mode flux is
\begin{equation}
\mathcal{F}_{22}=\frac{32}{5}\eta^2x^5\left|\widehat{h}_{22}\right|^2 .
\end{equation}
Energy conservation determines the time and accumulated GW phase,
\begin{equation}
\frac{dt}{dx}=-\frac{M}{\mathcal{F}_{22}}\frac{d(E_{\mathrm{real}}/M)}{dx}, \qquad
\frac{d\phi_{\mathrm{GW}}}{dx}=\frac{2x^{3/2}}{M}\frac{dt}{dx},
\end{equation}
where $E_{\mathrm{real}}$ is the total binary energy including the rest-mass contribution.

After integrating these equations, the frequency-domain waveform is
\begin{equation}
\widetilde{h}(f)=A(f)\exp\left\{i\left[2\pi f t_c-\phi_c-\frac{\pi}{4}+\Psi_{\mathrm{orbit}}(f)+\arg\widehat{h}_{22}(f)\right]\right\},
\end{equation}
with amplitude
\begin{equation}
A(f)=\sqrt{\frac{5}{24}}\frac{\mathcal{M}_c^{5/6}}{\pi^{2/3}d_L}f^{-7/6}
\sqrt{-\frac{2}{\eta}\frac{d(E_{\mathrm{real}}/M)}{dx}}\,W(f),
\end{equation}
where $W(f)$ smoothly suppresses the waveform near the last stable orbit.

The required artifact is a standalone Python implementation of
\begin{equation}
\widetilde{h}=\widetilde{h}\left(f;\mathcal{M}_c,\eta,d_L,t_c,\phi_c,\lambda_{\mathrm{RG}}\right),
\end{equation}
mapping an array of detector-frame frequencies to a complex waveform of the same size. The agent must consistently handle physical units, construct $x$, integrate the evolution equations, and implement the prescribed frequency cutoffs. The task therefore tests multi-stage formula-to-code reasoning: qualitatively plausible waveforms can still be scientifically incorrect because of unit inconsistencies, phase conventions, integration boundaries, or cutoff implementation.

Accuracy is measured using the frequency-domain mismatch $\mathcal{M}$ defined in Appendix~\ref{app:metrics}, evaluated over 144 combinations of chirp mass, symmetric mass ratio, luminosity distance, and $\lambda_{\mathrm{RG}}$. The validation set therefore tests whether the derived implementation remains accurate across both nearly equal-mass and unequal-mass binaries and across different strengths of the RG-resummed contribution.

%==============================================================================================
\subsubsection{Template Bank Bench}
\label{sec:template_bench}

Template banks are libraries of representative signals used to search noisy data. Similar ideas appear in radar, sonar, wireless communications, seismology, medical imaging, computer vision, spectroscopy, and gravitational-wave astronomy. In GW searches, a large collection of predicted signals is matched against detector data; nearly all black-hole merger signals in GW data have been found using such template-bank pipelines \cite{lvc_gwtc3_o3_ab_catalog_2021, Will24_Outside_LVK_catalog_review, Wad23_Pipeline, nitz_4ogc_o3_ab_catalog_2021}.
This benchmark evaluates whether agents can construct a compact template bank for frequency-domain GW signals~\citep{vandenbroeck2009templatebanks,schmidt2023templatebanks,Wad23_TemplateBanks,Cabass:2025wqg}. Unlike the waveform and dynamics benchmarks, which predict continuous outputs, this task is a coverage and optimization problem: we want to cover a given region of parameter space with the smallest set of representative waveforms to lower the matched-filtering computational cost while preserving accuracy.
We give agents a public pool of waveform parameters (corresponding to a particular high-mass black-hole parameter space) and allow them to call \texttt{LALSuite}\footnote{\href{https://lscsoft.docs.ligo.org/lalsuite/dev/index.html}{https://lscsoft.docs.ligo.org/lalsuite/dev/index.html}} routines~\citep{wette2020swiglal} to generate multimode, aligned-spin gravitational waveforms using the IMRPhenomXHM model~\citep{Garcia-Quiros:2020qpx}. The agent must then split the waveforms into training and validation sets, use the training set to construct an efficient template bank, and evaluate the bank using the validation waveforms. Each template is represented by parameters $\lambda = \{m_1, m_2, \chi_{1z}, \chi_{2z}, \phi_{\mathrm{ref}}\}$, where $m$, $\chi$ are the black-hole masses and spins and $\phi_{\mathrm{ref}}$ is a reference phase used to offset different modes in the waveform.
The objective is to achieve high coverage using as few templates as possible. We evaluate the performance by creating test waveforms hidden from the agent. Our test metric is the number of waveforms ($N_\mathrm{agent}$) from the output bank needed such that \(50\%\) of hidden-test waveforms achieve match \(\geq 0.97\) within the bank (see Appendix~\ref{app:metrics} for details). We compare the performance with reference solution $(N_\mathrm{domain})$ to achieve the same match using the mode-by-mode filtering approach of Refs.~\citep{Wad23_HM_Events,Wad23_Pipeline,Zhou:2026hcw,Cheung:2025grp, Mehta:2025jiq} (where the overlap from multiple modes is combined to form an optimal detection statistic). We show the relative efficiency $N_\mathrm{domain}/N_\mathrm{agent}$ in panel (g) of Fig.~\ref{fig:violin}.
This task differs from standard regression benchmarks because there is no direct predictor to fit: effective solutions must balance diversity and redundancy. Agents often use approaches similar to stochastic template placement, but fall short of the optimal waveform-decomposition approach~\citep{Wad23_TemplateBanks}.

%==============================================================================================
\section{Details of the Evaluation Metrics}
\label{app:metrics}
%==============================================================================================
This appendix defines the evaluation metrics used across the \texttt{gwBenchmarks} suite. We use notation consistent with Section~\ref{sec:gwbenchmarks}: $\lambda$ denotes the input parameter vector, and subscripts ``pred'' and ``ref'' indicate predicted and reference quantities. The separate evaluator calculates every metric using the same reference implementation for all agents. Equations~\eqref{eq:rem_err}, \eqref{eq:ring_err}, and~\eqref{eq:val_err} define dataset-level aggregates. Figure~\ref{fig:violin} instead displays the corresponding samplewise terms before the outer mean or root-mean-square aggregation, and Table~\ref{tab:main_results} reports their medians. This distributional summary exposes tails while the equations retain the prescribed dataset-level objectives used by the separate evaluator.

%----------------------------------------------------------------------------------------------
\subsection{Frequency-Domain Mismatch}
%----------------------------------------------------------------------------------------------

For waveform-based tasks (Sections~\ref{sec:waveform_bench}, \ref{sec:analytic_bench}, and \ref{sec:new_physics_bench}), model accuracy is evaluated using the standard frequency-domain mismatch. Given two waveforms $h_1$ and $h_2$, the mismatch is
\begin{equation}
\mathcal{M}(h_1,h_2)=1-\max_{t_0,\phi_0}\frac{\langle h_1,h_2\rangle}{\sqrt{\langle h_1,h_1\rangle\langle h_2,h_2\rangle}},
\label{eq:fd_mismatch}
\end{equation}
where the maximization is performed over relative time and phase shifts $(t_0,\phi_0)$. The noise-weighted inner product is
\begin{equation}
\langle h_1,h_2\rangle=4\,\mathrm{Re}\int_{f_{\mathrm{low}}}^{f_{\mathrm{high}}}\frac{\tilde{h}_1(f)\tilde{h}_2^*(f)}{S_n(f)}\,df,
\end{equation}
where $\tilde{h}(f)$ denotes the Fourier transform and $S_n(f)$ is the detector noise power spectral density.

All mismatch calculations use the \texttt{PyCBC} implementation~\citep{nitz2018pycbclive} with the \texttt{aLIGOZeroDetHighPower} PSD over $f\in[15,990]\,\mathrm{Hz}$. We report the mismatch averaged over total masses $M=m_1+m_2\in\{40,80,120,160,200\}\,M_\odot$. The mismatch lies in $[0,1]$, with $\mathcal{M}=0$ indicating identical waveforms. In GW modeling, high-accuracy models typically achieve mismatches $\lesssim10^{-3}$, with state-of-the-art models approaching $\sim10^{-4}$.

%----------------------------------------------------------------------------------------------
\subsection{Time-Series and Regression Errors}
%----------------------------------------------------------------------------------------------

For the Dynamics, Remnant, and Ringdown tasks (Sections~\ref{sec:dynamic_bench}, \ref{sec:remnant_bench}, and \ref{sec:ringdown_bench}), we use relative error measures adapted to the structure and scale of each problem.

For Dynamics (Section~\ref{sec:dynamic_bench}), the agent predicts a trajectory $x(t;\lambda)$ over a discrete time grid $\{t_i\}_{i=1}^{T}$. Accuracy is measured using the root-mean-squared relative error
\begin{equation}
\mathcal{L}_{\mathrm{dyn}}=\sqrt{\frac{1}{T}\sum_{i=1}^{T}\left[\frac{x_{\mathrm{pred}}(t_i;\lambda)-x_{\mathrm{ref}}(t_i;\lambda)}{x_{\mathrm{ref}}(t_i;\lambda)}\right]^2},
\label{eq:dyn_err}
\end{equation}
which measures relative deviations across the orbital evolution. The metric is non-negative and unbounded above. In practice, values $\lesssim10^{-3}$ indicate high-fidelity modeling, whereas values $\gtrsim10^{-2}$ correspond to significant deviations in the predicted evolution.

For Remnant (Section~\ref{sec:remnant_bench}), the task is to predict the recoil velocity $v_k(\lambda)$. We use the normalized root-mean-squared error
\begin{equation}
\mathcal{L}_{\mathrm{rem}}=\frac{\sqrt{\frac{1}{N}\sum_{i=1}^{N}\left[v_k^{\mathrm{pred}}(\lambda_i)-v_k^{\mathrm{ref}}(\lambda_i)\right]^2}}{\max(v_k^{\mathrm{ref}})-\min(v_k^{\mathrm{ref}})},
\label{eq:rem_err}
\end{equation}
where normalization by the reference range accounts for the wide dynamic range of recoil velocities. The metric is non-negative and typically $\ll1$. Values $\lesssim10^{-2}$ correspond to accurate predictions, while larger values indicate poorer generalization, often driven by difficult high-kick configurations.

For Ringdown (Section~\ref{sec:ringdown_bench}), the agent predicts complex quasi-normal-mode frequencies $\omega(\lambda)=\omega_r+i\omega_i$. Accuracy is measured using the mean relative error
\begin{equation}
\mathcal{L}_{\mathrm{ring}}=\frac{1}{N}\sum_{i=1}^{N}\frac{\left|\omega_{\mathrm{pred}}(\lambda_i)-\omega_{\mathrm{ref}}(\lambda_i)\right|}{\left|\omega_{\mathrm{ref}}(\lambda_i)\right|},
\label{eq:ring_err}
\end{equation}
which measures errors in both the oscillation and damping components of the complex frequency. The metric is non-negative and typically very small because the underlying mapping is smooth and low dimensional. High-quality models achieve errors $\lesssim10^{-8}$, with the best interpolation strategies approaching machine precision.

%----------------------------------------------------------------------------------------------
\subsection{Log-Space Error for Mismatch Prediction}
%----------------------------------------------------------------------------------------------

For Validity (Section~\ref{sec:validity_bench}), the agent predicts waveform mismatches $\hat{\mathcal{M}}(\lambda)$ spanning several orders of magnitude. To account for this dynamic range, accuracy is evaluated in log space:
\begin{equation}
\mathcal{L}_{\mathrm{val}}=\sqrt{\frac{1}{N}\sum_{i=1}^{N}\left[\log_{10}\hat{\mathcal{M}}_{\mathrm{pred}}(\lambda_i)-\log_{10}\hat{\mathcal{M}}_{\mathrm{ref}}(\lambda_i)\right]^2}.
\label{eq:val_err}
\end{equation}
This metric emphasizes relative rather than absolute accuracy across the error landscape. It is non-negative, with $\mathcal{L}_{\mathrm{val}}=0$ corresponding to exact predictions. Values $\lesssim0.1$ indicate good agreement across scales, whereas values $\gtrsim0.3$ reflect substantial discrepancies in the predicted mismatch landscape.

%----------------------------------------------------------------------------------------------
\subsection{Template-Bank Coverage}
%----------------------------------------------------------------------------------------------

For Template Bank (Section~\ref{sec:template_bench}), performance is defined by how efficiently an ordered set of waveform templates covers the hidden parameter space. Given a hidden-test waveform $h_i$ and the first $K$ templates $\{h_j\}_{j=1}^{K}$, the best overlap is
\begin{equation}
O_i(K)=\max_{j\leq K}\frac{\langle h_i,h_j\rangle}{\sqrt{\langle h_i,h_i\rangle\langle h_j,h_j\rangle}}.
\end{equation}
Waveforms are generated using standard \texttt{LALSuite} routines, and overlaps are computed using the same matched-filter inner product as in the mismatch calculation~\citep{allen2005findchirp}.

Performance is evaluated using the smallest prefix length $K_{50}$ for which at least $50\%$ of hidden-test waveforms satisfy
\begin{equation}
O_i(K_{50})\geq0.97.
\end{equation}
Thus, the benchmark measures how many templates are required to achieve the prescribed coverage rather than evaluating only the quality of the full submitted bank.

To compare agent-generated banks with the human-designed reference, we report the normalized efficiency
\begin{equation}
\mathrm{Efficiency}=\frac{K_{\mathrm{ref}}}{K_{50}},
\end{equation}
where $K_{\mathrm{ref}}=27$ is the size of the human-optimized template bank. Higher values indicate more efficient coverage, with $\mathrm{Efficiency}=1$ matching the reference performance and $\mathrm{Efficiency}>1$ indicating that the required coverage is achieved with fewer templates.

Individual overlaps lie in $[0,1]$, with $O_i=1$ indicating identical normalized waveforms. The threshold $O_i\geq0.97$ corresponds to a mismatch $\leq0.03$ and is a conventional criterion for sufficiently dense template-bank coverage in matched-filter GW searches.

%----------------------------------------------------------------------------------------------
\subsection{Evaluation Protocol}
%----------------------------------------------------------------------------------------------
As described in Section~\ref{sec:results}, the separate evaluator calculates all metrics on the complete validation sets using the same reference implementations and preprocessing for every agent. This prevents proxy metrics or partial evaluations from determining the reported results.

\end{document}